\documentclass[structabstract]{aa}  
\usepackage{graphicx}
\usepackage{txfonts}
\usepackage{natbib}
\bibpunct{(}{)}{;}{a}{}{,} % to follow the A&A style
\begin{document}
\title{Asteroseismology of hybrid $\delta$ Scuti--$\gamma$ Doradus 
pulsating stars} 
 
\author{J. P. S\'anchez Arias\inst{1,2}, A. H. C\'orsico\inst{1,2} \and   
        L. G. Althaus\inst{1,2}}
\institute{$^{1}$ Facultad de Ciencias Astron\'omicas y Geof\'{\i}sicas, 
          Universidad Nacional de La Plata, Paseo del Bosque s/n, 1900 
          La Plata, Argentina\\
          $^{2}$ Instituto de Astrof\'isica La Plata, CONICET-UNLP, Argentina\\
          \email{jsanchez,acorsico,althaus@fcaglp.unlp.edu.ar}     
           }
\date{Received ; accepted }

\abstract {Hybrid $\delta$ Scuti-$\gamma$ Doradus pulsating  stars
show acoustic  ($p$) oscillation modes typical of $\delta$ Scuti 
variable stars, and gravity ($g$) pulsation  modes characteristic 
of  $\gamma$ Doradus variable stars simultaneously excited. Observations 
from space missions like MOST, CoRoT, and \emph{Kepler}  have revealed 
a large number of hybrid  $\delta$ Scuti-$\gamma$ Doradus pulsators, 
thus paving the way for a exciting new channel for asteroseismic studies.}  
{We perform a detailed asteroseismological modeling of five hybrid  
$\delta$ Scuti-$\gamma$ Doradus stars.}{We employ a grid-based  modeling
approach to sound the internal structure of  the target stars by
employing a huge grid of  stellar models from the zero-age main sequence 
to the terminal-age main sequence, varying parameters  like stellar mass, 
effective temperature, metallicity and core overshooting. We compute their 
adiabatic radial ($\ell= 0$) and non-radial ($\ell= 1, 2, 3$) 
$p$ and $g$ mode periods. We employ two 
model-fitting procedures to searching for the models that best reproduce the
observed pulsation spectra of each target star, that is, the 
asteroseismological models.}{We derive the fundamental parameters 
and the evolutionary status of five hybrid $\delta$ Scuti-$\gamma$ 
Doradus variable stars recently observed with the CoRoT and \emph{Kepler} 
space missions: CoRoT 105733033, CoRoT 100866999, 
KIC 11145123, KIC 9244992, and HD 49434. The
asteroseismological model for each star results from
different criteria of model
selection, in which we take full advantage of the richness of periods
that characterizes the pulsation spectra of this kind of stars.}{} 
\titlerunning{Hybrid $\delta$ Sct--$\gamma$ Dor
stars} 
\maketitle
%
%________________________________________________________________

\section{Introduction}
\label{introduction}

At the present time, pulsating stars constitute one of the most
powerful tools for sounding the stellar interiors and to derive a
wealth of information about the physical structure and evolutionary
status of stars through asteroseismology \citep{2010aste.book.....A,
  2010csp..book.....B,2015pust.book.....C}.  Nowadays, a huge number
of variable stars are routinely discovered and scrutinized by space
missions like MOST \citep{2003PASP..115.1023W},  CoRoT
\citep{2009IAUS..253...71B}   and \emph{Kepler}
\citep{2010ApJ...713L..79K}, which include  long-term monitoring with
high-temporal resolution and high-photometric  sensibility for
hundreds of thousands  stars.  Among the most intensively studied
classes of variable stars  in recent years we found the $\delta$ Scuti
(Sct) and  $\gamma$ Doradus (Dor), which are $\sim 1.2-2.2 M_{\sun}$
stars with spectral types between A and F, undergoing quiescent core H
burning at (or near of) the Main  Sequence (MS) ($ 6500\ {\rm K}
\lesssim T_{\rm eff} \lesssim 8500$ K).  They exhibit multiperiodic
brightness variations due to global radial and non-radial pulsation
modes.  

The $\delta$ Sct stars, discovered  over a century ago
\citep{1900ApJ....12..254C}, display high-frequency variations with
typical periods in the   range $\sim 0.008\ {\rm d} - 0.42$ d and
amplitudes from milli-magnitudes  up to almost one magnitude in blue
bands. They are likely produced by  nonradial $p$ modes of low radial
order $n$ and low harmonic degree  ($\ell= 1-3$),  although the
largest amplitude variations probably are induced by the radial
fundamental mode ($n= 0, \ell= 0$) and/or  low-overtone radial
modes ($n= 1,2,3, \cdots, \ell= 0$). The fact that $\delta$ Sct 
stars pulsate in nonradial  $p$
modes and radial modes implies that they are potentially useful  for
probing the stellar envelope.  The $\delta$ Sct
variables are  Population I  stars\footnote{There exist a Population
  II counterpart to $\delta$ Sct variables, the so called SX Phoenicis
  (Phe), which are usually observed in low-metallicity globular clusters
  \citep[see, e.g.,][]{2014RMxAA..50..307A}.} of  spectral type
between A0 and F5, lying on the extension  of the Cepheid instability
strip towards low luminosities, at  effective temperatures between
7000 K and 8500 K, stellar masses in the interval  $1.5-2.2 M_{\sun}$,
and luminosities in the range $5 \lesssim L/L_{\sun} \lesssim 80$
\citep{2015pust.book.....C}. The  projected  rotation velocities are
in the range $[0,150]$ km s$^{-1}$, although  they can reach values up to
$\sim 250$ km s$^{-1}$. The pulsations  are thought to be driven by
the $\kappa$ mechanism
\citep{1980tsp..book.....C,1989nos..book.....U}  operating in the
partial ionization zone of He II
\citep{1971A&A....14...24C,2004A&A...414L..17D,2005A&A...434.1055G}.
Solar-like oscillations stochastically driven have also been predicted to
occur in $\delta$ Sct stars \citep{2002A&A...395..563S}. Notably,
these  expectations  have been confirmed in one object
\citep{2011Natur.477..570A}. Among $\delta$ Sct stars,  frequently a
distinction is made between the so called  \emph{high-amplitude
  $\delta$ Sct stars} (HADS), whose amplitudes  in the V band exceed
0.3 mag, and their much more abundant  \emph{low-amplitude  $\delta$
  Sct stars} (LADS) counterparts  \citep[see][]{2008PASJ...60..551L}. 

The $\gamma$ Dor variables  \citep{1999PASP..111..840K}, were
recognized as a new class of pulsating stars about 20 years ago
\citep{1994MNRAS.270..905B}.  They are generally cooler than  $\delta$
Sct stars, with $T_{\rm eff}$ between 6700 K and 7400 K  (spectral
types between A7 and F5) and masses in the range  $1.5-1.8 M_{\sun}$
\citep{2015pust.book.....C}.  The $\gamma$ Dor stars pulsate in
low-degree, high-order $g$ modes  driven by a flux modulation
mechanism (``convective blocking'') induced  by the outer convective
zone \citep{2000ApJ...542L..57G, 2004A&A...414L..17D,
  2005A&A...434.1055G}. The low-frequency variations shown by these
stars  have periods typically between $\sim 0.3$ d and $\sim 3$ d and
amplitudes below $\sim 0.1$ magnitudes.  The presence of $g$ modes in
$\gamma$ Dor stars offers the chance of probing into the core regions.  
In addition, since high-order $g$ modes are
excited ($n \gg 1$), it is possible to use the  asymptotic theory
\citep{1980ApJS...43..469T} and the departures  from uniform period
spacing  (by mode trapping) to explore the possible chemical
inhomogeneities  in the structure of the convective cores
\citep{Miglio2008}.  Stochastic excitation of solar like oscillations
has been also predicted in $\gamma$ Dor stars
\citep{2007A&A...464..659P}, but no positive detection has been
reported yet. 

The instability strips of  $\delta$ Sct and $\gamma$ Dor stars
partially overlaps in the Hertzprung-Russell (HR) diagram  \citep[see,
  for instance, Fig. 4 of][]{2013A&A...556A..52T}, strongly suggesting
the existence of $\delta$ Sct-$\gamma$ Dor \emph{hybrid} stars,  that
is, stars showing high-frequency $p$-mode pulsations  typical of
$\delta$ Sct stars simultaneously with low-frequency  $g$-mode
oscillations characteristic of  $\gamma$ Dor stars
\citep{2004A&A...414L..17D, 2010ApJ...713L.192G}. The first  example
of a star pulsating intrinsically with both  $\delta$ Sct and $\gamma$
Dor frequencies was detected from  the ground
\citep{2005AJ....129.2026H}. Other examples are HD 49434
\citep{2008A&A...489.1213U} and HD 8801 \citep{2009MNRAS.398.1339H}. A
large sample of \emph{Kepler} and  CoRoT stars yielded the first hints
that  hybrid behavior might be common  in A-F type stars
\citep{2010ApJ...713L.192G, 2010arXiv1007.3176H}.  A follow up study
with a large ($> 750$ stars) sample  of $\delta$ Sct  and $\gamma$ Dor
candidates by \citet{2011A&A...534A.125U} revealed  that out of 471
stars showing $\delta$ Sct or $\gamma$ Dor pulsations, 36\% (171
stars) are hybrid $\delta$ Sct-$\gamma$ Dor stars.  Very recent
studies \citep[e.g.,][]{2015AJ....149...68B}  analyzing larger samples
of $\delta$ Sct or $\gamma$ Dor candidates strongly suggest that
hybrid $\delta$ Sct-$\gamma$ Dor stars are very
common. \citet{2015MNRAS.452.3073B} studied  the frequency
distributions of $\delta$ Sct stars observed  by the \emph{Kepler}
telescope in short-cadence mode and found low frequencies (typical of
$\gamma$ Dor stars)  in \emph{all} the analyzed $\delta$ Sct
stars. This finding  renders somewhat meaningless the concept of
$\delta$ Sct-$\gamma$ Dor hybrids. 

Apart from these important investigations of large samples of stars,
there are published studied on several individual  hybrid $\delta$
Sct-$\gamma$ Dor stars observed from  space missions. Among them, we
mention HD 114839  \citep{2006CoAst.148...28K} and  BD+18-4914
\citep{2006CoAst.148...34R}, both detected  by the MOST satellite.  On
the other hand, hybrid $\delta$ Sct-$\gamma$ Dor stars  discovered
with CoRoT observations are CoRoT 102699796
\citep{2011MNRAS.416.1535R},  CoRoT 105733033
\citep{2012A&A...540A.117C},  CoRoT 100866999
\citep{2013A&A...556A..87C},  and HD 49434
\citet{2015MNRAS.447.2970B}.  Finally, among hybrid stars discovered
with the \emph{Kepler} mission,  we mention KIC 6761539
\citep{2012AN....333.1077H},  KIC 11145123
\citep{2014MNRAS.444..102K},  KIC 8569819 \citep{2015MNRAS.446.1223K},
KIC 9244992 \citep{2015MNRAS.447.3264S}, KIC 9533489
\citep{2015A&A...581A..77B}, KIC 10080943 \citep{2015MNRAS.454.1792K}.

A few attempts of asteroseismic modeling of $\delta$ Sct  stars have
proven to be a very difficult task
\citep{2001AJ....122.2042C,2003A&A...411..503C,2008A&A...478..855L,
  2013MNRAS.432.2284M}. In part, this is due to that generally there
are many combinations of the stellar structure parameters ($T_{\rm
  eff}, M_{\star}, Y, Z$, overshooting, etc) that lead to very
different seismic  solutions but that reproduces with virtually the
same  degree of precision the set of observed frequencies. The
situation is potentially much more favorable in the case of hybrid
$\delta$ Sct-$\gamma$ Dor stars, because the simultaneous presence of
both $g$ and $p$ nonradial  modes (in addition to radial pulsations)
excited, that allows to place strong constraints on the whole
structure,  thus eliminating  most of the degeneration of
solutions. As such,  hybrid stars  have a formidable
asteroseismological potential and are very attractive targets for
modeling.  

Among the above mentioned hybrid objects, detailed  seismological
modeling  has so far been performed only for a few $\delta$
Sct-$\gamma$ Dor hybrids, namely KIC 11145123
\citep{2014MNRAS.444..102K}, KIC 9244992 \citep{2015MNRAS.447.3264S},
and the binary system  KIC 10080943 \citep{2016arXiv160507958S}. In
this study, we present  a seismic modeling of five $\delta$
Sct-$\gamma$ Dor hybrids including those two studied by
\citet{2014MNRAS.444..102K} and \citet{2015MNRAS.447.3264S}. Our
approach consists in the  comparison of the observed pulsation periods
with the theoretical adiabatic  pulsation periods (and period spacings
in the case of high-order $g$  modes)  computed on a huge set of
stellar models representative  of  A-F MS stars with masses in the
range $1.2-2.2 M_{\sun}$ generated   with a state-of-the-art
evolutionary code. This approach is frequently referred to as
\emph{grid-based} or \emph{forward} modeling in the field of
solar-like oscillations
\citep[e.g.,][]{2011ApJ...730...63G,2014A&A...564A.105H} and has been
the preferred asteroseismological approach in pulsating white dwarfs
\citep{2008A&A...478..869C,2010A&ARv..18..471A,2012MNRAS.420.1462R}.
Furthermore, this approach has been adopted for the study of  $\delta$
Sct-$\gamma$ Dor hybrid stars by \citet{2016arXiv160507958S}.  The
characteristics of the target stars are determined by  searching among
the grid of models to get a ``best-fit model'' for  a given observed
set of periods of radial modes, and $p$ and $g$  nonradial modes. In
particular, we make full use of the valuable  property that some hybrid
stars offer, that is, the value of the  mean period spacing
$\left(\overline{\Delta \Pi}\right)$ of $g$ modes. Specifically,  we
will perform an asteroseismic modeling of the hybrid $\delta$
Sct-$\gamma$ Dor stars CoRoT 105733033, \citep{2012A&A...540A.117C},
CoRoT 100866999, \citep{2013A&A...556A..87C},   KIC 11145123
\citep{2014MNRAS.444..102K},  KIC 9244992 \citep{2015MNRAS.447.3264S},
and  HD 49434 \citet{2015MNRAS.447.2970B}. The use of
$\overline{\Delta \Pi}$  allows to discard a large portion of the grid
of models  ---those models that do not reproduce the observed period
spacing. Also we assume, as usual, that the largest amplitude  mode in
the $\delta$ Sct region of the pulsation spectrum is  associated to
the fundamental radial mode ($\ell= 0, n= 0$) or the first radial
overtone modes ($\ell= 0, n= 1,2,3,4,\cdots$).  This step further reduces
the number of possible seismological  models. Finally, we perform a
period-to-period fit to the  $p$ mode periods. We also carry out other
possible  model selections, for instance, by performing direct
period-to-period fits to the complete set of observed periods
(including individual periods of $g$ modes and radial modes).  This
research constitutes the first stage of an ongoing systematic
asteroseismic modeling program of hybrid $\delta$ Sct-$\gamma$ Dor
stars  in the La Plata Observatory.

The paper is organized as follows: in Sect. \ref{tools} we describe
our evolutionary and pulsation numerical tools. The main ingredients 
of the model grid we use to assess the pulsation properties of
 hybrid $\delta$ Sct-$\gamma$ Dor stars are described in 
Sect. \ref{Modelling}. Sect. \ref{impact} is devoted to describe in 
the effects that core overshooting and metallicity have
on the pulsation properties of $g$ and $p$ modes.
In Sect. \ref{applications} we present our 
asteroseismic analysis of the target stars in detail. 
Finally, in Sect. \ref{conclusions} we summarize our main findings.

\section{Numerical tools}
\label{tools}

\subsection{Stellar evolution code}
\label{lpcode}

We have carried out an asteroseismological analysis of $\delta$
Sct-$\gamma$ Dor stars by computing a huge grid of evolutionary
and pulsational models representative of this kind of variable 
stars. The complete grid of models and some of their relevant
pulsation properties will be described in detail in
Sect. \ref{Modelling}. The stellar models were generated with the
help of the {\tt LPCODE} \citep{Althausetal2005} stellar evolution 
code which has been developed entirely at La Plata Observatory. 
{\tt LPCODE} is a well tested and widely
employed stellar code which is able to simulate the evolution of 
low- and intermediate-mass stars from
the zero-age main sequence (ZAMS), through the core H-burning phase, 
the He-burning phase, and the thermally pulsing asymptotic giant branch 
(AGB) phase to the white dwarf (WD) stage. The code has been used in a 
variety of studies involving the formation and evolution 
of WDs \citep{Renedo2010,Althaus2012,2013A&A...555A..96S}, 
extremely low-mass (ELM) WD
stars \citep{Althausetal2013}, H-deficient PG1159 stars
resulting from very late thermal pulses (VLTP) and the 
``born-again scenario'' \citep{Althausetal2005, Miller2006}, 
sdB and sdO stars \citep{Miller2008, 2012ASPC..452..175M}, and low-mass 
giant stars considering fingering convection 
\citep{Wachlin2011,2014A&A...570A..58W}. 

{\tt LPCODE} is based on the \citet{Kippenhahn1967} method for
calculating stellar evolution. The code has the capability  to generate
stellar models with an arbitrary number of mesh points by means of an
algorithm that adds mesh points where they are needed ---where
physical variables change appreciably---, and eliminates them where
they are not necessary. The main physical ingredients of {\tt LPCODE},
relevant for our analysis  of hybrid $\delta$ Sct-$\gamma$ Dor,
include: radiative  opacities of OPAL project
\citet{IglesisasyRogers1996}, complemented at low temperatures with
the molecular opacities produced by \citet{Ferguson2005}; equation of
state (EoS) at low-density regime of OPAL project, comprising the
partial ionization for H and He compositions, the radiation pressure
and the ionic contribution; the nuclear network considers the
following 16 elements: $^{1}$H, $^{2}$H, $^{3}$He, $^{4}$He, $^{7}$Li,
$^{7}$Be, $^{12}$C, $^{13}$C, $^{14}$N, $^{15}$N, $^{16}$O, $^{17}$O,
$^{18}$O, $^{19}$Fe, $^{20}$Ne, $^{22}$Ne and 34 thermonuclear
reaction rates to describe the H (proton-proton chain and CNO
bi-cycle) and He burning and C ignition. 
The abundance changes for all chemical elements are described 
by the set of equations:

\begin{equation}
\left(\frac{dY}{dt}\right)=
\left(\frac{\partial Y}{\partial t}\right)_{\rm nuc} + 
\frac{\partial}{\partial M_r}\left[(4\pi r^2 \rho)^2 
D\frac{\partial Y}{\partial M_r}\right],
\label{difusion}
\end{equation}

\noindent being $Y$, the vector containing fractions of all
the considered nuclear species. The first 
term of Eq. (\ref{difusion}) gives the abundance changes due
to thermonuclear reactions. Details about the numerical procedure for
solving these equations can be found in \citet{Althaus2003}. All of
ours models have a convective core, since we considered masses in the
range $1.2-2.2M_{\sun}$. For some of them we
consider the occurrence of core overshooting, 
that is, mixing of chemical elements beyond the formal convective 
boundary which is set by the Schwarzschild
criterion $\nabla_{\rm ad}< \nabla_{\rm rad}$ 
\citep[$\nabla_{\rm ad}$ and $\nabla_{\rm rad}$ being the adiabatic 
and radiative temperature gradients; see][]{Kippenhahn2012} 
\footnote{Semiconvection, that is, the mixing of layers for which 
$\nabla_{\rm ad} < \nabla_{\rm rad} < \nabla_{\rm L}$ 
\citep[where $\nabla_{\rm L}$ is the Ledoux 
temperature gradient; see][for its definition]{Kippenhahn2012}
 is not taken into account. We note that 
semiconvective layers on top of the convective zone may
vary the extent of the convective core.}. In {\tt LPCODE}, 
the mixing due to convection, salt finger\footnote{The
salt finger instability takes place
when the stabilizing agent (heat) diffuses away faster than the
destabilizing agent ($\mu$), leading to a slow mixing process that
might provide extra mixing \citep{2007A&A...467L..15C}.}and overshoot 
is treated as a diffusion process (second term of 
Eq. \ref{difusion}). The efficiency of
convective and salt-finger mixing is described by appropriate
diffusion coefficients $D$ which are specified by our treatment of
convection. Here, we adopted the classical mixing length theory (MLT)
for convection \citep[see, e.g.,][]{Kippenhahn2012} with the free
parameter $\alpha= 1.66$, with which we reproduce the present
luminosity and effective temperature of the Sun, $L_{\sun}= 3.842
\times 10^{33}\ {\rm erg\ s}^{-1}$ and $\log T_{\rm eff}= 3.7641$, when
$Z= 0.0164$ and $X= 0.714$ are adopted, according to the $Z/X$ value of
\citet{Grevessenoels1993}. Extra-mixing episodes
(overshooting) are taken into account  as time dependent 
diffusion process, assuming that the mixing velocities  decay exponentially
beyond convective boundaries with a diffusion coefficient given by: 

\begin{equation}
D= D_0 \exp(-2 z / f H_{\rm p})
\label{efe}
\end{equation}

\noindent where $D_0$ is the diffusive coefficient near the edge of
the convective zone, $z$ is the geometric distance of the considered
layer to this edge, $H_{\rm P}$ is the pressure scale height at the
convective boundary, and $f$ is a measure of the extent of the overshoot
region \citep{Herwig1997, Herwig2000}. In this study we
consider several values for $f$ (see Sect. 
\ref{Modelling}).

\subsection{Pulsation code}
\label{lp-pul}

The pulsation computations employed in this work were carried out
with the adiabatic version of the {\tt LP-PUL} pulsation  code
described in detail in \citet{2006A&A...454..863C},  which is coupled
to the {\tt LPCODE} evolutionary code.  Briefly, the {\tt LP-PUL} 
pulsation code
is based on the general Newton-Raphson technique to solve the full set
of equations and boundary conditions that describe linear, adiabatic,
radial and non-radial stellar pulsations, following the dimensionless
formulation of \citet{1971AcA....21..289D}. 
The pulsation code provides the dimensionless eigenfrequency $\omega_n$ 
---$n$ being the radial order of the mode--- and eigenfunctions 
$y_1, \cdots, y_4$. From these basic quantities, the code
computes the pulsation periods ($\Pi_n$), the oscillation kinetic 
energy ($K_n$), the first order rotation splitting coefficients 
($C_n$), the weight functions ($w_n$), and the variational periods
($\Pi^{\rm v}_n$) for each computed eigenmode. Generally, the 
relative difference between $\Pi^{\rm v}_n$ and $\Pi_n$ is lower 
than $\sim 10^{-4}$ (that is, $\sim 0.01 \%$). This represents the 
precision with which {\tt LP-PUL} code computes the pulsation periods 
\citep[see][for details]{2002Ap&SS.279..281C}. The set of pulsation
equations, boundary conditions, and pulsation quantities of 
relevance to this work are  given in Appendix section of \citet{2006A&A...454..863C}. The {\tt LP-PUL} pulsation 
code has been tested and extensively  used in numerous 
asteroseismological studies of pulsating WDs and pre-WDs 
\citep{2001A&A...380L..17C,2005A&A...429..277C,2006A&A...454..863C,2008A&A...478..869C, 2012MNRAS.424.2792C,2014ApJ...793L..17C,2014A&A...569A.106C, 2012MNRAS.420.1462R,2013ApJ...779...58R},  as well as hot subdwarf stars 
sdB and sdO \citep{2011ApJ...741L...3M, 2012ASPC..452..175M}. 
Recently, it has been employed for the first time to explore the pulsation 
properties of $\delta$ Sct and $\gamma$ Dor stars \citep{Sanchezetal2013}.

For  $g$ modes  with  high   radial  order  $n$ (long  periods),  the
separation  of consecutive  periods ($|\Delta  n|= 1$)  becomes nearly
constant  at a  value  dependent on $\ell$, given  by the  asymptotic  
theory of  non-radial
stellar  pulsations. In {\tt LP-PUL}, the asymptotic period spacing for 
$g$ modes is computed as in \citet{Tassouletal1990}:

\begin{equation}
\Delta \Pi^{a}_{\ell}= \frac{2\pi^2}{\sqrt{\ell(\ell+1)}} 
\left[\int_{r_1}^{r_2}\frac{N}{r}dr\right]^{-1} 
\label{aps}
\end{equation}

\noindent where $r_1$ and $r_2$ are the radius of the inner and outer
boundaries of the propagation region, respectively. For  $p$ modes  
with  high   radial  order  (high  frequencies),  the
separation  of consecutive  frequencies  becomes nearly
constant and independent on $\ell$,  at a  value given by 
\citep{1989nos..book.....U}:

\begin{equation} 
\Delta \nu^{\rm a} = \left[ 2 \int_0^R \frac{dr}{c_{\rm s}}\right]^{-1},
\label{afs}
\end{equation}

\noindent where $c_{\rm s}$ is the local adiabatic sound speed, 
defined as $c_{\rm s}^2 =\Gamma_1 P / \rho$.

In our code, the squared Lamb frequency ($L_{\ell}$, one of the  
critical frequencies of nonradial stellar pulsations) is computed 
 as:

\begin{equation} 
L_{\ell}^2 = \ell (\ell+1) \frac{c_{\rm s}^2}{r^2}
\label{lamb}
\end{equation}

On the other hand, the squared Brunt-V\"ais\"al\"a frequency 
($N$, the other critical frequency of nonradial stellar pulsations) 
is computed as \citep{Tassouletal1990}:  

\begin{equation}
\label{bvf}
N^2= \frac{g^2 \rho}{P}\frac{\chi_{\rm T}}{\chi_{\rho}}
\left[\nabla_{\rm ad}- \nabla + B\right],
\label{bv}
\end{equation}

\noindent where the compressibilities are defined as

\begin{equation}
\chi_{\rho}= \left(\frac{\partial\ln P}{\partial \ln \rho}\right)_{{\rm T}, 
\{\rm X_i\}}\ \ \
\chi_{\rm T}= \left(\frac{\partial \ln P}{\partial \ln T}\right)_{\rho, 
\{\rm X_i\}}.
\end{equation}

\noindent The Ledoux term $B$ is computed as:

\begin{equation}
\label{B}
B= -\frac{1}{\chi_{\rm T}} \sum_1^{M-1} \chi_{\rm X_i} \frac{d\ln X_i}{d\ln P}, 
\label{BLedoux}
\end{equation}

\noindent where:

\begin{equation}
\chi_{\rm X_i}= \left(\frac{\partial \ln P}{\partial \ln X_i}\right)_{\rho, 
{\rm T}, \{\rm X_{j \neq i}\}}.
\end{equation}

\noindent The explicit contribution of a chemical composition 
gradient to the Brunt-V\"ais\"al\"a frequency is contained in term 
$B$. This formulation of the Brunt-V\"ais\"al\"a frequency, which is 
particularly suited for WD stars \citep{1991ApJ...367..601B},
can be reduced to the usual expression of $N^2$ in presence of 
varying composition \citep{1980tsp..book.....C,Miglio2008}. If we assume 
a nondegenerate and completely ionized ideal gas,
which is valid in the deep interiors of MS stars like $\delta$ Sct 
and $\gamma$ Dor stars, Eq. (\ref{bv}) reduces to: 

\begin{equation}
N^2= \frac{g^2 \rho}{P} 
\left[\nabla_{\rm ad}- \nabla + \nabla_{\mu} \right],
\label{bv-MS}
\end{equation}

\noindent where 

\begin{equation}
\nabla_{\mu}= \frac{d\ln \mu}{d\ln P},
\end{equation}

\noindent $\mu$ being the mean molecular weight.

\section{Model grid and pulsation computations}
\label{Modelling}

The stellar models used in this paper were calculated from the  ZAMS
up to the stages in which the H abundance  at the core is negligible
($X_{\rm H}\lesssim 10^{-6}$),  defining the Terminal Age Main
Sequence  (TAMS). The initial Hydrogen abundance ($X_{\rm H}$) adopted
in the ZAMS varies according to the selected metallicity. 

 In the present analysis, we considered stellar masses  between $1.2$
 and $2.2M_{\sun}$ with a mass step of  $\Delta M_{\star}=
 0.05M_{\sun}$. This mass interval embraces the  range of masses
 expected for most of the $\delta$ Sct  and $\gamma$ Dor stars. We
 consider three different values for the metallicity to generate our
 evolutionary sequences: $Z= 0.01, 0.015, 0.02$, thus, as it was
 mentioned before, the initial H abundance ($X_{\rm H}$) adopted in
 the ZAMS varies according to the selected metallicity through the
 relation between the metallicity and the initial He abundance
 ($Y_{\rm He}$): $Y_{\rm He}= 0.245 + 2 \times Z$ and $X_{\rm H} +
 Y_{\rm He} + Z= 1$.

 In addition, we have taken into account the occurrence 
of extra-mixing episodes in the form of convective overshooting. Since 
overshooting is poorly constrained, we adopted four 
different cases: no overshooting ($f= 0$), moderate 
overshooting ($f= 0.01$), intermediate overshooting 
($f= 0.02$), and extreme overshooting ($f= 0.03$) 
(see Eq. \ref{efe} for the definition of $f$). By varying all
these parameters we computed a large set of $21 \times 3 \times 4= 252$ 
evolutionary sequences.  To have a dense grid of stellar models 
for each sequence ---thus allowing us to carefully follow the 
evolution of the internal structure and also the pulsation properties
of the models---
we fixed the time step of {\tt LPCODE} to have stellar 
models that differs $\sim 10-20$ K in $T_{\rm eff}$.
Thus, each evolutionary sequence computed from the ZAMS to the TAMS 
comprises $\sim 1500$ models. All in all, we have computed 
$\sim 400000$ stellar models that constitute a sufficiently 
dense and comprehensive grid of equilibrium models representative 
of  hybrid $\delta$ Sct-$\gamma$ Dor variable stars
as to ensure a consistent search for a seismological model
for each target star. 

For each equilibrium  model we have computed adiabatic radial 
($\ell= 0$) and nonradial ($\ell= 1, 2, 3$) $p$ and $g$ modes
with pulsation periods in the range 
$0.014\ {\rm d} \lesssim \Pi_n \lesssim 3.74$ d 
($1200\ {\rm sec} \lesssim \Pi_n \lesssim 300000$ sec), 
thus amply embracing the range of periods usually detected 
in hybrid $\delta$ Sct-$\gamma$ Dor stars.  
Models were divided into approximately $1400$ mesh points and 
their distribution was regularly updated every 1 evolutionary 
time-step. The number of mesh points proved to be high enough as
to solve the rapidly oscillating eigenfunctions of  
high-radial order modes smoothly.

\begin{figure}[h!] 
\begin{center}
\includegraphics[clip,width=8.5 cm]{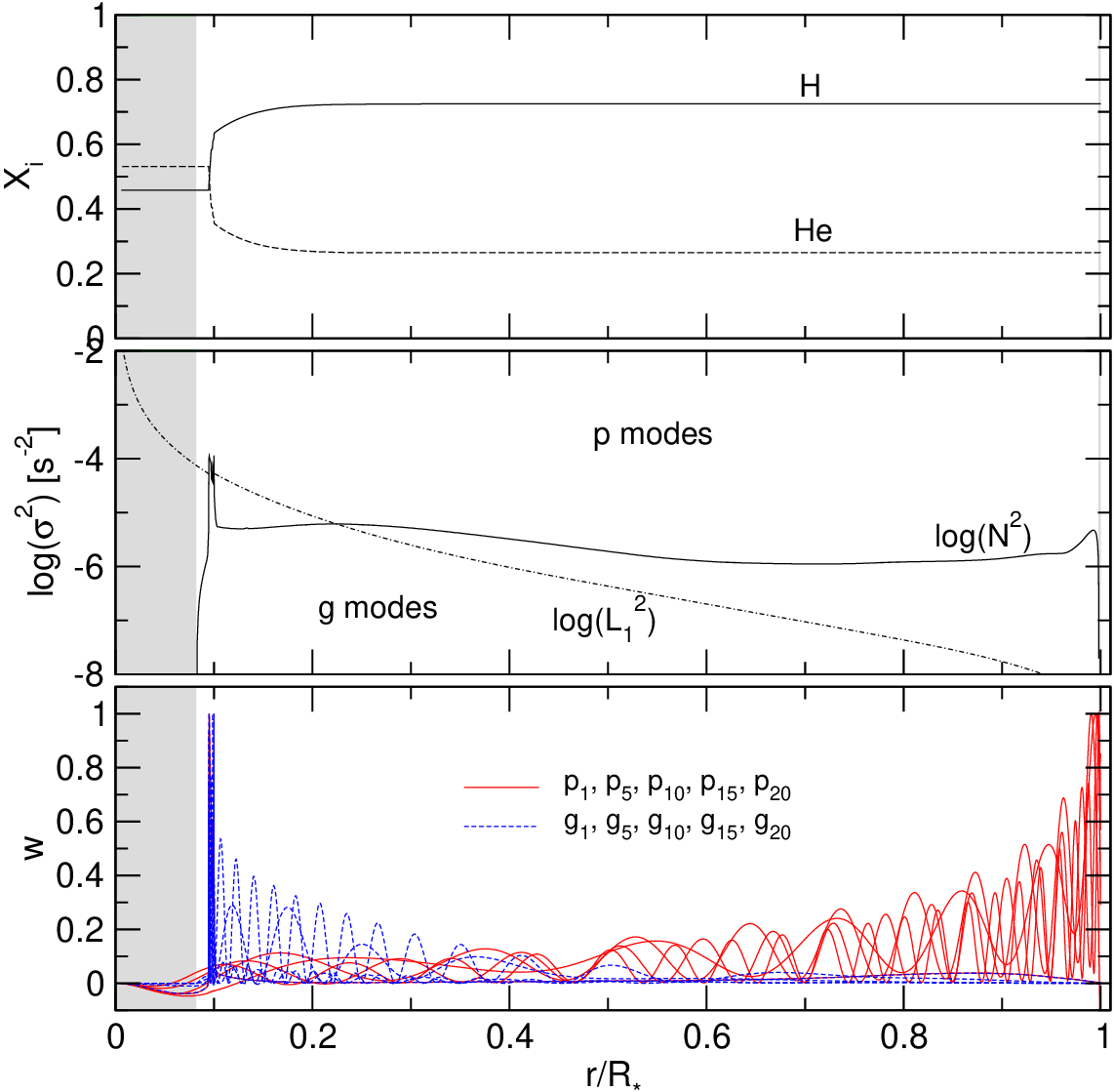} 
\caption{H and He fractional abundances (upper panel), 
the logarithm of the squared Lamb and Brunt-V\"ais\"al\"a frequencies, 
computed according to Eqs. (\ref{lamb}) and (\ref{bv}), respectively 
(central panel), and the weight function (Eq. (A.14) of the Appendix of \citet{2006A&A...454..863C}) of 
dipole ($\ell= 1$) $p$- and $g$-modes with 
radial order $n= 1, 5, 10, 15, 20$ (lower panel), corresponding to a 
typical MS stellar model with $M_{\star}= 1.5 M_{\sun}$, 
$L_{\star}= 8.34 L_{\sun}$, 
$Z= 0.01$, $f= 0.01$ and $\tau= 1.06$ Gyr.}
\label{hybrids} 
\end{center}
\end{figure}

\begin{figure}[h!] 
\begin{center}
\includegraphics[clip,width=9 cm]{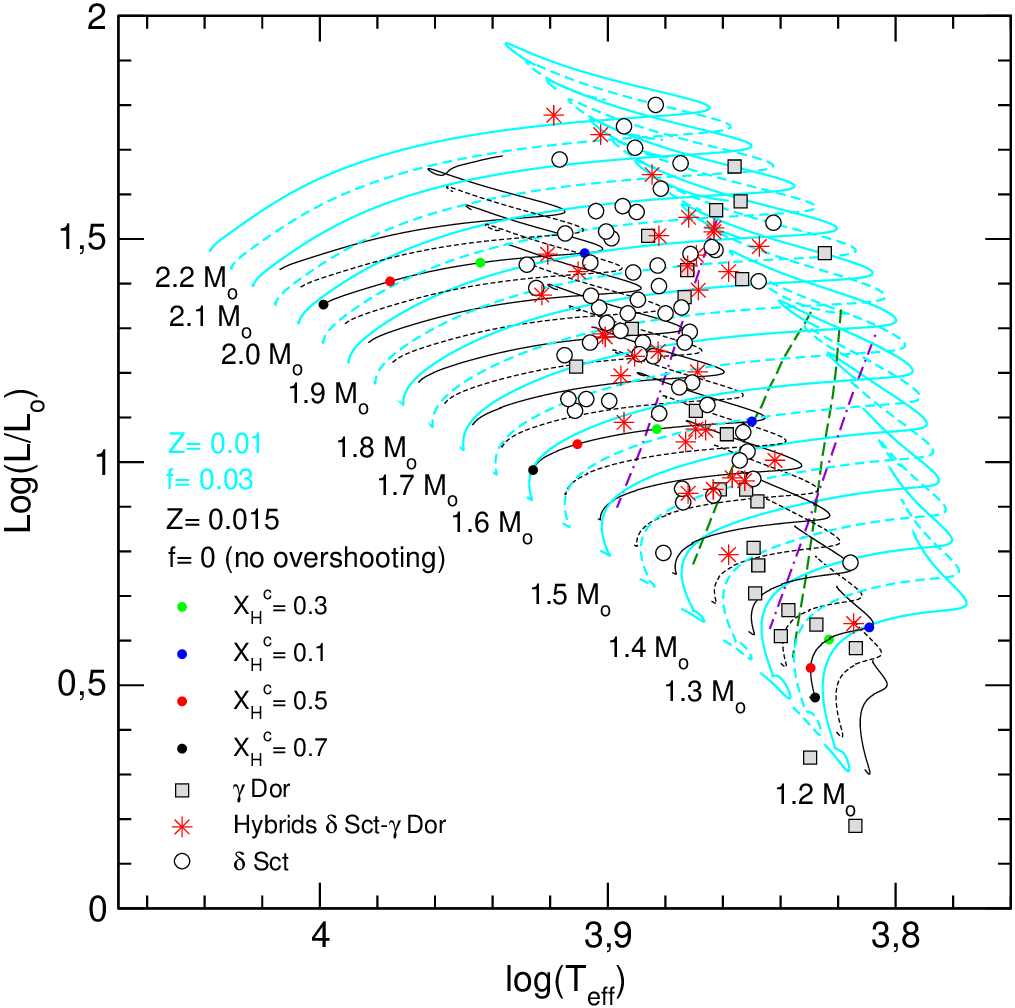} 
\caption{HR diagram showing evolutionary tracks for stellar models
with different masses ($1.2 \leq M_{\star}/M_{\sun}\leq 2.2$), 
$Z= 0.015$ and without overshooting ($f= 0$) in black, and $Z=0.01$ and $f=0.03$ in light blue, from the ZAMS to the TAMS.
The value of the stellar mass ($M_{\star}$) is indicated 
for a subset of the tracks (those displayed with solid lines).
Black, red, green, and blue dots correspond to the 
location of stellar models with $M_{\star}/M_{\sun}= 1.3, 1.7$ and $2.1$  
having a central H abundance of $X_{\rm H}^{\rm c}= 0.7, 0.5, 0.3, 0.1$, 
respectively. A sample of $\delta$ Sct (open circles), 
$\gamma$ Dor (gray squares), and hybrid 
$\delta$ Sct-$\gamma$ Dor (red star symbols) stars taken 
from \citet[][]{2010ApJ...713L.192G} are included for illustrative
purposes. Also, the boundaries of the $\delta$ Sct (violet dot-dashed lines) 
and $\gamma$ Dor (green dashed lines) theoretical instability 
strips from \citet{2005A&A...435..927D} are plotted.}
\label{figure-HRM} 
\end{center}
\end{figure}

Next, we describe some properties of our stellar models. We choose, 
in particular, a template stellar model with $M_{\star}= 1.5 M_{\sun}$, 
$L_{\star}= 8.34 L_{\sun}$, $Z= 0.01$, $f= 0.01$.
This model is burning H at the core, and has a stellar age
of $\tau= 1.06$ Gyr. In Fig. \ref{hybrids}, we depict some characteristics
of this model. Specifically, the upper panel displays
the fractional abundances of H and He in terms of normalized 
radius  ($r/R_{\star}$). For this particular model, 
the central H and He abundances are $0.458$ and $0.531$, respectively. The 
model is characterized by a convective core from the stellar 
center ($r/R_{\star}= 0$) to a radius $r/R_{\star} \sim 0.082$, 
which is emphasized with a gray area. Note that, due 
to overshooting, H and He are mixed up to
a location somewhat beyond the boundary of the convective core, 
at $r/R_{\star}\sim 0.095$. The model 
has also a very thin outer convection zone, barely visible in the figure, 
that extends from $r/R_{\star}\sim 0.998$ to the surface ($r/R_{\star}= 1$).
The middle panel of Fig. \ref{hybrids} depicts a ``propagation 
diagram'' \citep[see][]{1980tsp..book.....C,1989nos..book.....U,
2015pust.book.....C}, that is, a diagram in 
which the Brunt-V\"ais\"al\"a and the Lamb frequencies (or the squares
of them) are plotted versus the stellar radius (or other similar 
coordinate). A local analysis of the pulsation equations
assuming the so-called Cowling  approximation ($\Phi'= 0$) and
high-order modes ($n\gg 1$), shows that $p$ and $g$ modes follow 
a dispersion relation given by:

\begin{equation}
k_r^2= \sigma^{-2} c^{-2} \left(\sigma^2 - N^2
\right) \left( \sigma^2 -L_{\ell}^2\right), 
\end{equation}

\noindent which relates the local radial wave 
number $k_r$  to the pulsation  frequency $\sigma$ 
\citep{1989nos..book.....U}. We note that if  
$\sigma^2  >  N^2, L_{\ell}^2$  or  $\sigma^2  <  N^2,
L_{\ell}^2$, the wave  number $k_r$ is real, and if  $N^2 > \sigma^2 >
L_{\ell}^2$ or $N^2 < \sigma^2 < L_{\ell}^2$, $k_r$ is purely imaginary. 
As  we can see in the  figure, there exist two
propagation regions,  one corresponding to  the case $\sigma^2  > N^2,
L_{\ell}^2$, associated with  $p$-modes, and other in which
the   eigenfrequencies  satisfies   $\sigma^2   <  N^2,   L_{\ell}^2$,
associated with  $g$-modes. 
Finally, in the lower panel of Fig. \ref{hybrids},
we display the normalized weight functions 
---defined by Eq. (A.14) of the Appendix of \citet{2006A&A...454..863C}--- corresponding 
to $\ell= 1$ nonradial $p$ and $g$ modes with 
radial orders $n= 1, 5, 10, 15, 20$.  
The weight functions indicate the regions of  the star that most 
contribute  to the period formation \citep{1985ApJ...295..547K}.
The figure shows very clearly that $p$ modes are relevant 
for probing the outer stellar regions, and $g$ modes are essential for 
sounding the deep core regions of the star. It is precisely  
this property that renders hybrid $\delta$ Sct-$\gamma$ Dor stars 
as exceptional targets for asteroseismology.

In Fig. \ref{figure-HRM} we show a HR diagram displaying a subset of 
evolutionary sequences computed for this work. They correspond
to stellar masses from $1.2 M_{\sun}$ to $2.2 M_{\sun}$, 
$Z= 0.015$, and no overshooting (\textbf{$f= 0$}) in black, and 
$Z= 0.01$ and $f= 0.03$ in light blue. The Figure includes the 
evolutionary stages comprised between the ZAMS and the TAMS.
Some $\delta$ Sct, $\gamma$ Dor, and hybrid $\delta$ Sct-$\gamma$ 
Dor stars \citep[taken from][]{2010ApJ...713L.192G}\footnote{Other 
larger and recent samples can be found in \citet{2015AJ....149...68B,
2015MNRAS.452.3073B}.}, along with the blue and red edges of 
the $\delta$ Sct and $\gamma$ Dor theoretical instability domains  
according to \citet{2005A&A...435..927D}, are included for 
illustrative purposes. 

We emphasize that, in the sake of simplicity, the impact of stellar rotation on the equilibrium models and on the pulsation spectra has been neglected in this work. Admittedly, this simplification could not be entirely valid for $\delta$ Scuti/$\gamma$ Doradus stars. As a matter of fact, even at moderate rotation, the rotational splitting may completely (or almost completely) destroy the regularities in the spectra, in particular, the period spacing between consecutive high-order gravity modes of non-rotating models \citep[see, for instance, Fig. 13 of][]{1993MNRAS.265..588D} for the case of SPB model stars. We defer to a future publication a comprehensive study of the impact of rotation on the period spectrum of $\delta$ Scuti/$\gamma$ Doradus stars.

\section{The impact of extra mixing and metallicity on the pulsation 
properties}
\label{impact}

\begin{figure}[h!] 
\begin{center}
\includegraphics[clip,width=8.5 cm]{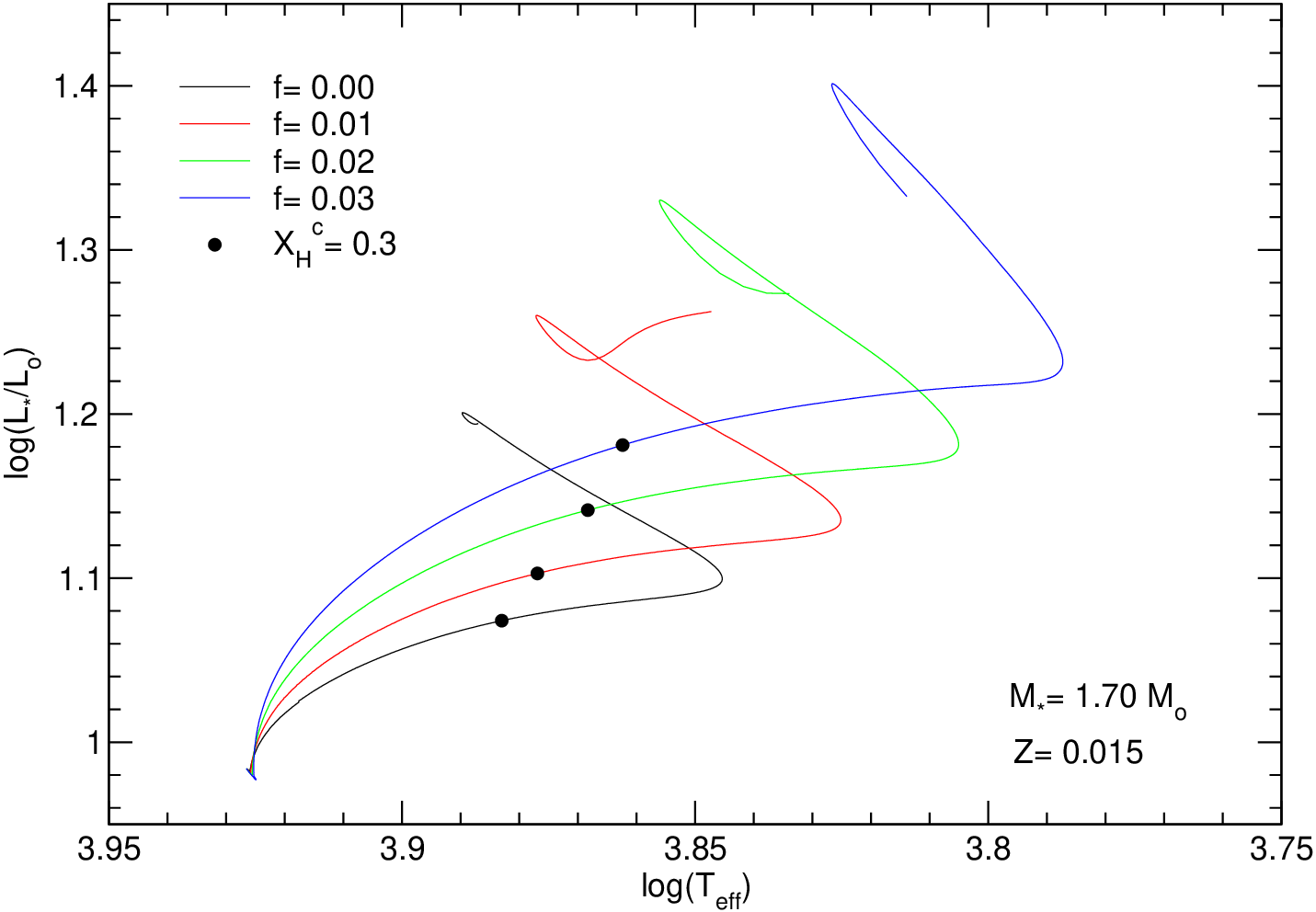} 
\caption{HR diagram showing the evolutionary tracks of models with 
$M_{\star}= 1.70 M_{\sun}$, $Z= 0.015$ and different overshooting
parameters ($f= 0.00, 0.01, 0.02, 0.03$). Selected models having 
a central H abundance of  $X_{\rm H}\sim  0.3$ are marked along the tracks 
with black dots.}
\label{figure-HROV} 
\end{center}
\end{figure}

Below,  we illustrate the effects of varying the amount of core 
overshooting (measured by $f$) and metallicity ($Z$) on our evolutionary 
models. The impact of 
core overshooting for different values of $f$ on the shape and extension of 
evolutionary tracks are shown in the HR diagram of Fig. \ref{figure-HROV}
for the case of stellar models with $M_{\star}= 1.70 M_{\sun}$
and $Z= 0.015$. As it can be seen, the occurrence of overshooting 
extends the incursion of the model star towards higher luminosities and 
lower effective temperatures, resulting ultimately in a notable 
broadening of the MS. This is because extra mixing promotes 
the existence of a larger amount of H available for burning 
in the stellar core. The lifetimes of stars in the MS are 
increased accordingly. Note that overshooting do not change
the location of the ZAMS in the HR diagram, which for this particular 
example, has $\log T_{\rm eff}\sim 3.925 $ and 
$\log(L_{\star}/L_{\sun})\sim 0.98$. 

\begin{figure}[h!] 
\begin{center}
\includegraphics[clip,width=8.5 cm]{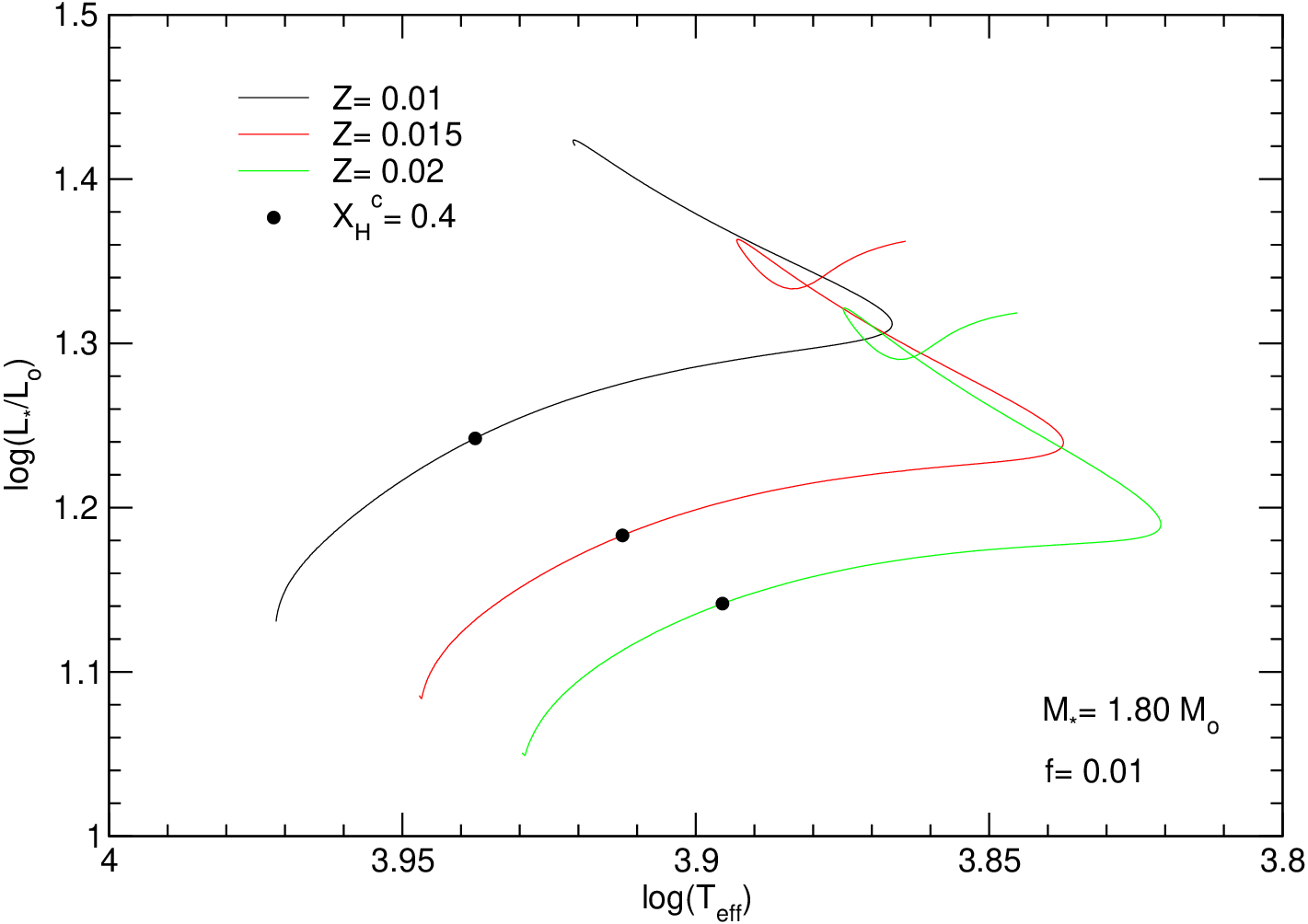} 
\caption{HR diagram showing evolutionary tracks of models with 
$M_{\star}= 1.80 M_{\sun}$, $f= 0.01$ and different metallicities
($Z= 0.01, 0.015$, and 0.02). Selected models having 
a central H abundance of  $X_{\rm H}\sim  0.4$ are marked along the tracks 
with black dots.}
\label{figure-HRZ} 
\end{center}
\end{figure}

The effects of metallicity on the evolutionary tracks is depicted in
Fig. \ref{figure-HRZ}, where we show a HR diagram for model sequences
computed assuming different metallicities ($Z= 0.01, 0.015$ and 0.02)
with $M_{\star}= 1.80 M_{\sun}$ and $f= 0.01$. It is apparent that, at
variance with the effect of overshooting, when we change $Z$ the
location  of the ZAMS is notoriously affected. Indeed, reducing the
metallicity  from $Z= 0.02$ to $Z= 0.015$ increases ZAMS effective
temperature  from $\log T_{\rm eff}\sim 3.93$ to $\log T_{\rm eff}\sim
3.95$  ($0.02$ dex), and the ZAMS luminosity from
$\log(L_{\star}/L_{\sun})\sim 1.05$ to $\log(L_{\star}/L_{\sun}) \sim
1.08$ ($0.03$ dex). In summary, reducing the  metallicity results in
hotter and more luminous models.  This can be understood on the basis
that, in spite of the fact that low-$Z$ models experience some 
reduction in the CNO cycle luminosity, this dimming is compensated 
by the reduction of the opacity at the photosphere, which results 
in bluer and brighter stars \citep[see, e.g.,][]{2004sipp.book.....H,
2005essp.book.....S}.

\begin{figure}[h!]
\begin{center}
\includegraphics[clip,width=9 cm]{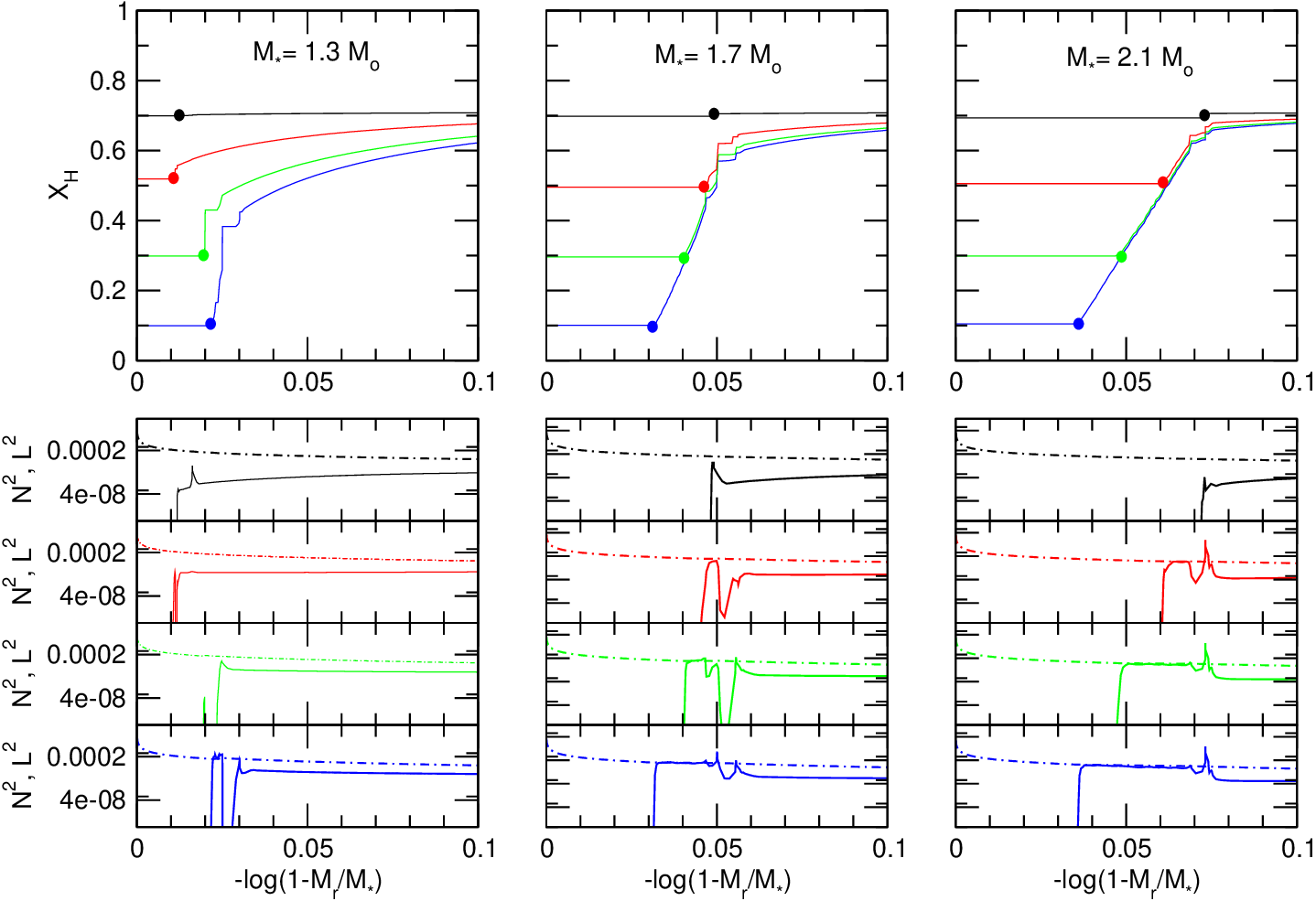} 
\caption{The H abundance profile (upper panels) and
the squared Brunt-V\"ais\"al\"a (full lines) and Lamb 
(dashed lines) frequencies (lower panels) for $M_{\star}= 1.30 M_{\sun}$ (left), 
$M_{\star}= 1.70 M_{\sun}$ (center), and 
$M_{\star}= 2.10 M_{\sun}$ (right).
 The four different 
evolutionary states displayed are clearly distinguishable
from the different central abundances of 
H ($X_{\rm H}= 0.7, 0.5, 0.3, 0.1$). The models, which were computed
with $Z= 0.015$ and $f= 0.00$, are marked in Fig. \ref{figure-HRM} 
with colored dots.} 
\label{figure-profilesM} 
\end{center}
\end{figure}

We turn now to show some $p$- and $g$-mode pulsational properties 
of our models. \citet{Miglio2008} have thoroughly investigated the 
properties of 
high-order $g$ modes for stellar models with masses in the range 
$1-10 M_{\sun}$ in the MS and the effects of stellar mass, hydrogen abundance at the core and extra-mixing processes on the period spacing features. 
So, we will frequently invoke results of \citet{Miglio2008} 
to compare with our own results. 
In Fig. \ref{figure-profilesM} we 
show the H chemical profile ($X_{\rm H}$) at the core regions (in terms 
of the mass fraction coordinate $-\log(1-M_r/M_{\star})$), 
associated to different evolutionary stages at the MS (upper panels),
and the respective squared Lamb and Brunt-V\"ais\"al\"a frequency runs 
(lower panels) corresponding to stellar models with masses 
$M_{\star}= 1.30 M_{\sun}$ (left) $M_{\star}= 1.70 M_{\sun}$ (center), 
and $M_{\star}= 2.10 M_{\sun}$ (right). The models were computed 
with $Z= 0.015$  and disregarding core overshooting. 
The location of these models is shown in Fig. \ref{figure-HRM} 
with colored dots. In the upper panels of Fig. \ref{figure-profilesM},
the boundary of the convective core for each evolutionary stage 
is marked with a dot. The four evolutionary stages shown 
correspond to central H abundances of $X_{\rm H}= 0.7, 0.5, 0.3$ 
and $0.1$. For the model with $M_{\star}= 1.30 M_{\sun}$ (left upper panel of 
Fig. \ref{figure-profilesM}), we obtain a 
growing convective core, that is, the mass of the
convective core increases during part of the MS 
\citep[compare with Figures 11 and 14 of][]{Miglio2008}. As
the convective core gradually grows, a discontinuity 
in the chemical composition occurs at its boundary, 
as can be appreciated from the figure.

The situation is  markedly different for the models with $M_{\star}=
1.70 M_{\sun}$  and $M_{\star}= 2.10 M_{\sun}$ (center and right upper
panels of  Fig. \ref{figure-profilesM}), characterized by a receding
convective core \citep[compare with Figure 15 of][]{Miglio2008}.  In
this case, the mass of the  convective core shrink during part of the
evolution at the MS.  No chemical discontinuity is formed in this
situation,  although a chemical gradient is left at the edge of the
convective  core. The situation is qualitatively similar for both, the
$1.70 M_{\sun}$ and the $2.10 M_{\sun}$  models, being  a slightly
larger size of the convective core (for a fixed central H abundance)
for the most massive model the only difference.  

The impact of the chemical gradient  at the boundary of the convective
core on the Brunt-V\"ais\"al\"a frequency is apparent from the lower
panels of Fig. \ref{figure-profilesM}. The specific contribution of
the H/He chemical transition to $N$ is entirely contained in the term
$\nabla_{\mu}$ in Eq. (\ref{bv-MS}) (or, alternatively in the term $B$
in Eq. \ref{bv}). The feature in $N$ induced by the chemical gradient
is very narrow for the lowest-mass  model shown ($M_{\star}= 1.30
M_{\sun}$) due to the abrupt change at the boundary of the convective
core characterizing this model.  This feature becomes more extended
for more massive models  ($M_{\star}= 1.70 M_{\sun}$ and $M_{\star}=
2.10 M_{\sun}$) in  response to the less steep H/He chemical
transition region at the  edge of the convective core.

\begin{figure}[h!]
\begin{center}
\includegraphics[clip,width= 9 cm]{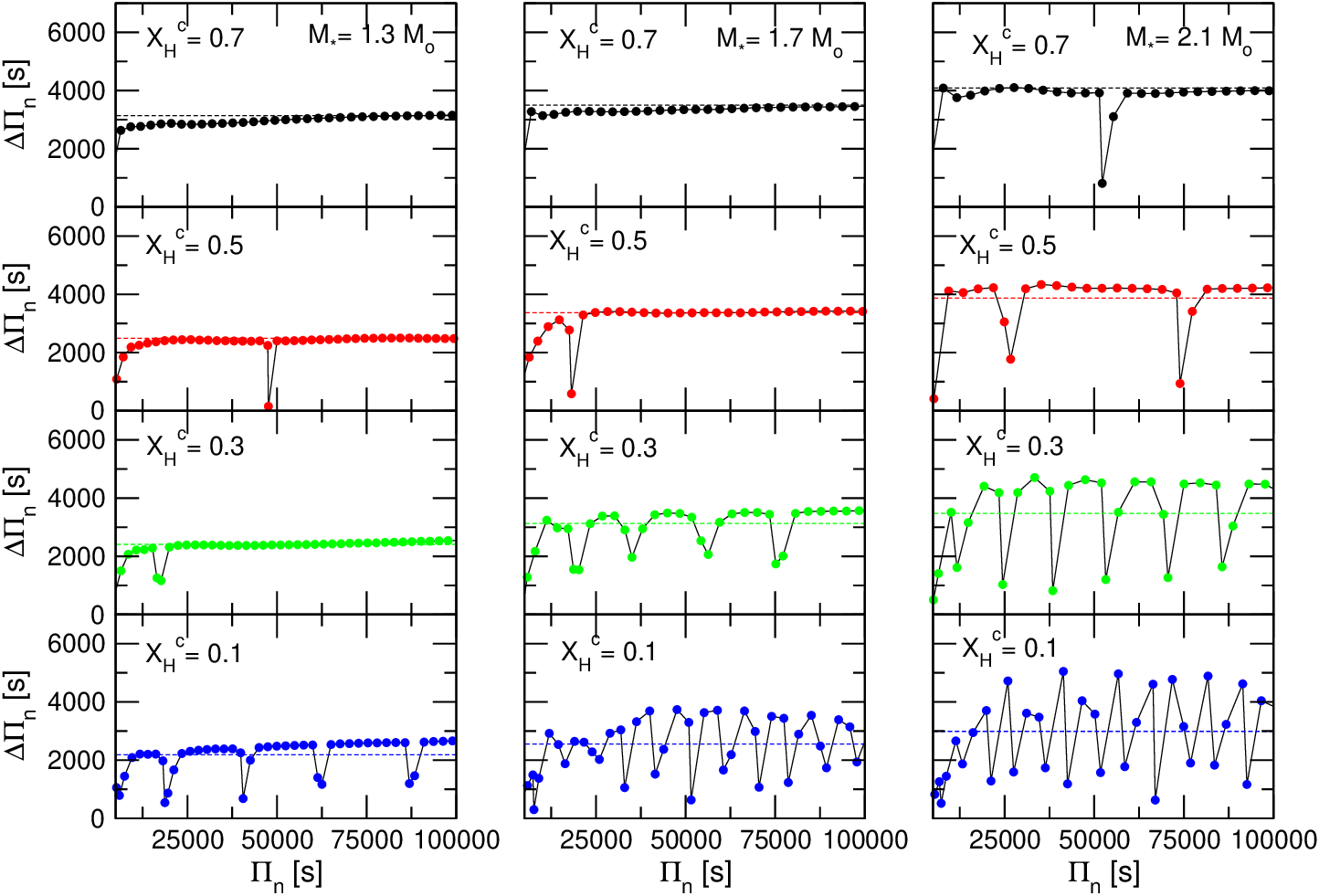} 
\caption{The dipole ($\ell= 1$) forward period spacing 
($\Delta \Pi_n$) of $g$ modes in terms of the periods $\Pi_n$ 
corresponding to the same stellar models with $M_{\star}= 1.30 M_{\sun}$ 
(left), $M_{\star}= 1.70 M_{\sun}$ (middle) and $M_{\star}= 2.10 M_{\sun}$ 
(right) shown in Fig. \ref{figure-profilesM}. The horizontal 
thin dashed lines correspond to the asymptotic period spacing 
($\Delta \Pi^{a}_{\ell= 1}$) computed according to 
Eq. (\ref{aps}). The H abundance at the stellar 
centre ($X_{\rm H}^{\rm c}$) is indicated in each panel.}
\label{figure-DPvPM} 
\end{center}
\end{figure} 

The presence of a chemical composition gradient in the interior of 
a star has a strong impact on the spacing of $g$-mode periods 
with  consecutive radial order, much like what happens in white 
dwarf pulsators that gives place to the resonance 
phenomena called \emph{mode trapping}
\citep{1992ApJS...80..369B,Corsico2002}.  Indeed, \citet{Miglio2008}
have shown in detail how the deviations from a constant period spacing
can yield information on the chemical  composition gradient left by a
convective core. In Fig. \ref{figure-DPvPM} we show the 
$\ell= 1$ forward period spacing of $g$ modes, defined as 
$\Delta \Pi_n= \Pi_{n+1}-\Pi_n$, in terms of the pulsation periods, 
$\Pi_n$, corresponding to the same stellar models 
shown in Fig. \ref{figure-profilesM}. We include in the plots
the asymptotic period spacing, $\Delta \Pi^{a}_{\ell= 1}$, depicted with thin horizontal dashed lines. For a fixed abundance of H 
at the core ($X_{\rm H}^{\rm c}$), the asymptotic period spacing 
(and thus the mean period spacing) increases with increasing 
stellar mass, as predicted by  Eq. (\ref{aps}). In fact, for 
more massive models the size of the convective core is larger, 
leading to a smaller range of integration in the integral of
Eq. (\ref{aps}). Thus, the integral is smaller 
and so $\Delta \Pi^{a}_{\ell= 1}$ is larger.  For a fixed 
stellar mass,  on the other hand, the asymptotic period 
spacing decreases with age (lower $X_{\rm H}^{\rm c}$ values),
because the integral increases when more evolved models are considered
\citep[see Figs. 14 and 15 of][]{Miglio2008}.

Fig. \ref{figure-DPvPM} dramatically shows how the period 
spacing is affected by 
the presence of the composition gradient, resulting in multiple
minima of $\Delta \Pi_n$, whose number increases as 
the star evolves on the MS ($X_{\rm H}^{\rm c}$ 
diminishes). For instance, for $M_{\star}= 2.10 M_{\sun}$ 
(right panels of Fig. \ref{figure-profilesM})
and $X_{\rm H}^{\rm c}= 0.7$, the model is just leaving the ZAMS, 
and there is a small step in the H profile (barely visible in the plot), 
which results in a peak in the Brunt-V\"ais\"al\"a frequency. 
The resulting period spacing is mostly constant, except 
for the presence of a strong minimum in $\Delta \Pi_n$
(upper right panel in Fig. \ref{figure-DPvPM}). As the star evolves 
and gradually consumes the H in the core, the feature 
in the Brunt-V\"ais\"al\"a frequency widens, and the edge of the 
convective core moves inward. When $X_{\rm H}^{\rm c}= 0.5$,
the forward period spacing exhibits two strong minima. At the 
stage in which $X_{\rm H}^{\rm c}= 0.3$, $\Delta \Pi_n$
is no longer constant, but shows numerous minima (six in the 
period range displayed in the Figure). This trend is further 
emphasized when the star is almost reaching the TAMS  
($X_{\rm H}^{\rm c}= 0.1$), as clearly depicted in the lowest
right panel of Fig. \ref{figure-DPvPM}. Our results are qualitatively 
similar to those of \citet{Miglio2008} 
(see their Figures 14 and 15). In particular, these authors
have derived explicit expressions that provide the periodicity
(in terms of $n$) of the oscillatory component in the period 
spacing $\Delta \Pi_n$ and its connection with the spatial location of the 
sharp variation in $N$ caused by the chemical gradient. 

\begin{figure}[h!]
\begin{center}
\includegraphics[clip,width= 9 cm]{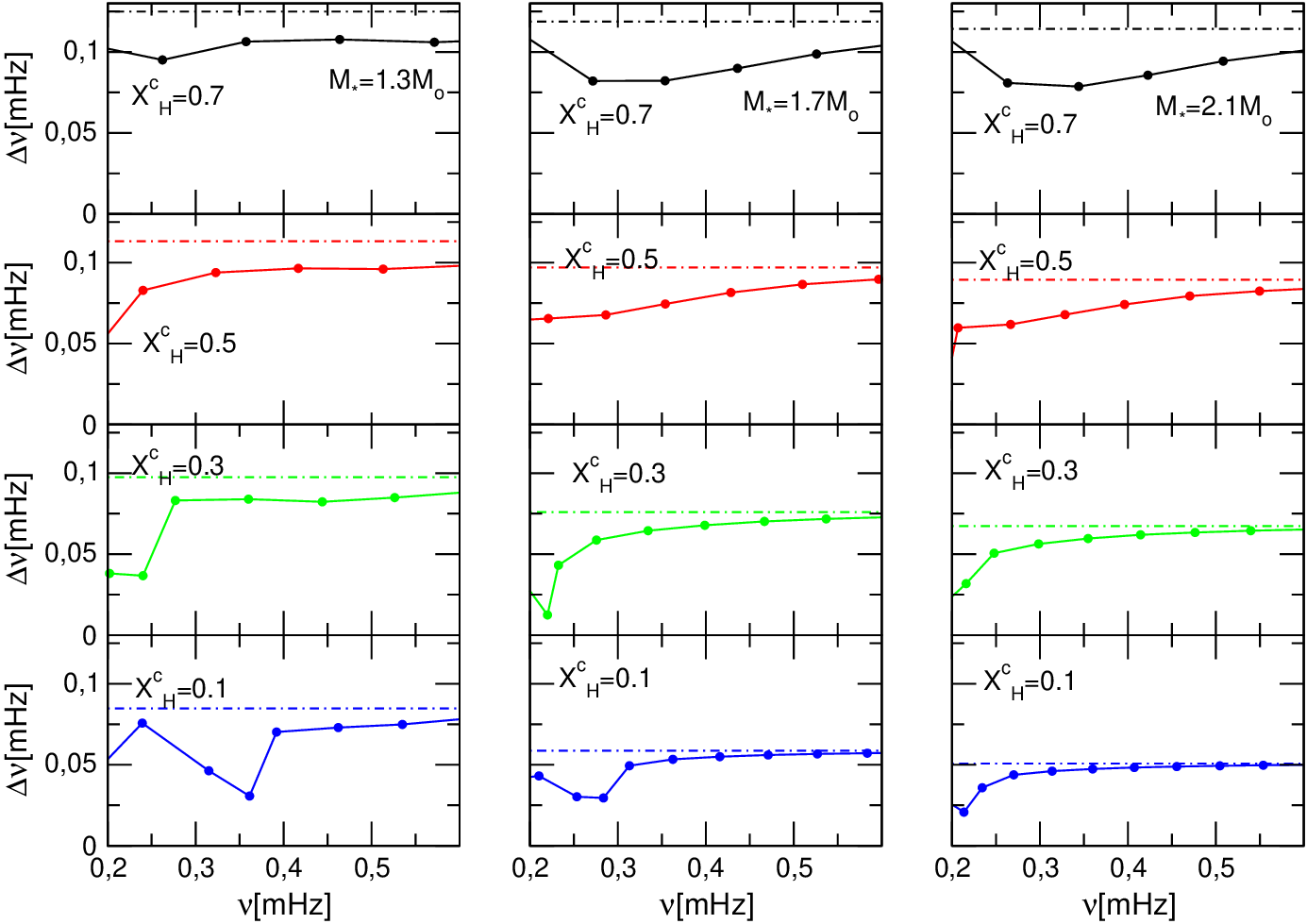} 
\caption{The dipole ($\ell= 1$) forward frequency spacing 
($\Delta \nu $) of $p$ modes in terms of the frequencies $\nu$ 
corresponding to the same stellar models with $M_{\star}= 1.30 M_{\sun}$ 
(left), $M_{\star}= 1.70 M_{\sun}$ (middle) and $M_{\star}= 2.10 M_{\sun}$ 
(right) shown in Fig. \ref{figure-profilesM}. The horizontal 
thin dashed lines correspond to the asymptotic frequency spacing 
($\Delta \nu^{a}_{\ell= 1}$) computed according to 
Eq. (\ref{afs}). The H abundance at the stellar 
center ($X_{\rm H}^{\rm c}$) is indicated in each panel.}
\label{figure-DnvnM} 
\end{center}
\end{figure}

We also investigate the impact (if any) of the chemical gradient on
the Lamb frequency which is depicted in lower  panels of
Fig. \ref{figure-profilesM} with dashed lines. There is no apparent
influence of this gradient on the Lamb frequency. However, we found
that $L_l^{2}$ exhibit a little bump (not visible in the plot) due to
the chemical gradient. On the other hand, the Lamb frequency is lower
for massive models. The forward frequency spacing for p modes
corresponding to models shown in Fig. \ref{figure-profilesM} is
depicted in Fig. \ref{figure-DnvnM} for the different masses
considered. In this plot, it can be seen that the asymptotic frequency
spacing and $\Delta \nu$ decrease with the evolution (i.e. lower
$X_{\rm H}$ values) for each considered mass.  
 
\begin{figure}[h!] 
\begin{center}
\includegraphics[clip,width=9 cm]{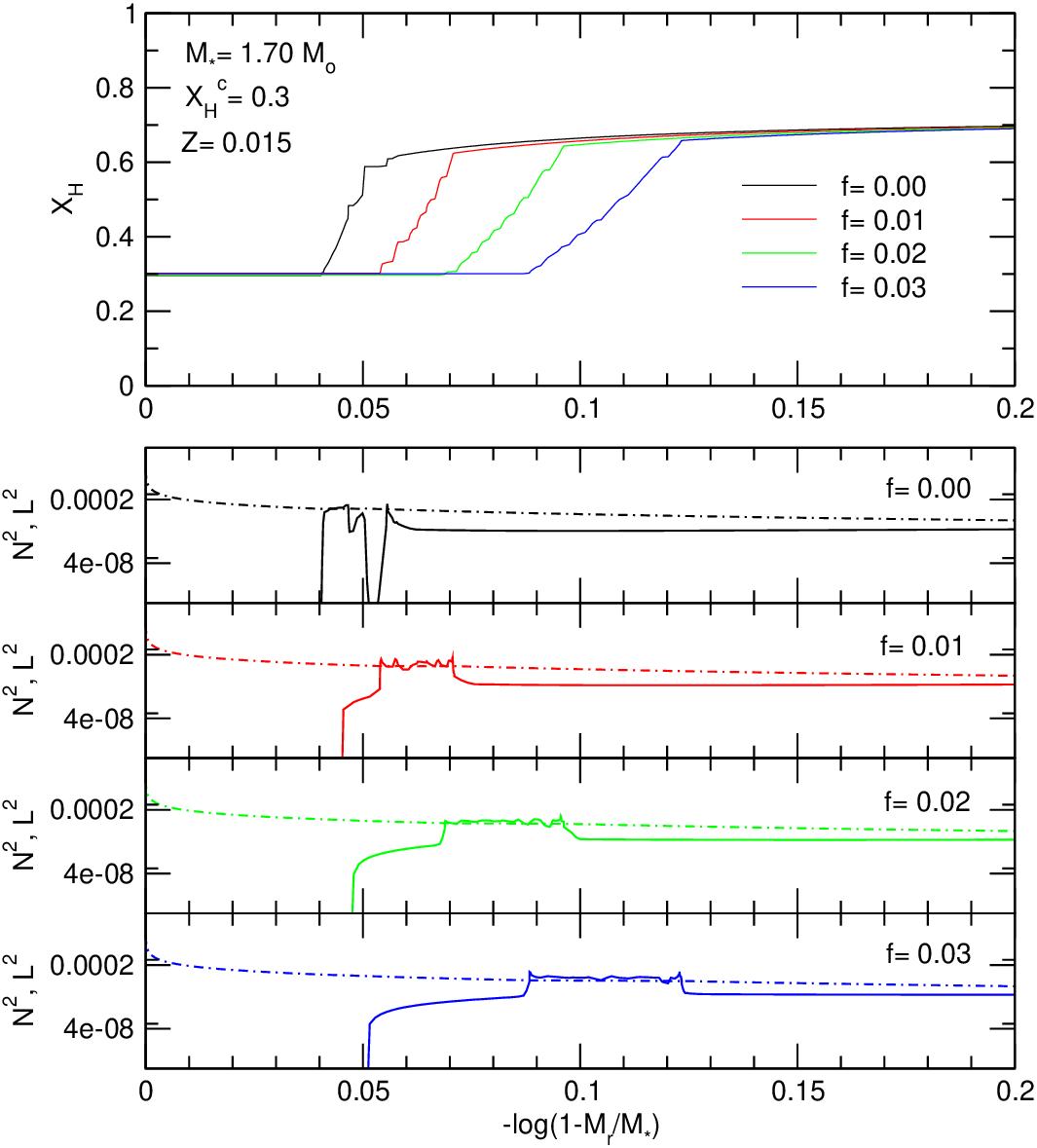} 
\caption{H abundance (upper panel), and the logarithm of the 
squared Brunt-V\"ais\"al\"a and Lamb frequencies (lower panels) corresponding 
to the stellar models with $M_{\star}= 1.7 M_{\sun}$, 
$Z= 0.015$, $X^{\rm c}_{\rm H}= 0.3$ (central abundance)
and different assumptions for the value of $f$ of the 
convective overshooting. The location of these models in the HR diagram are 
marked in Fig. \ref{figure-HROV} with black dots.} 
\label{figure-rvxhOV} 
\end{center}
\end{figure}

Now, we describe the effects of core overshooting on the $p$- and $g$-mode 
pulsational properties of our models. In the upper panel
of Fig. \ref{figure-rvxhOV} we depict the H abundance 
corresponding to stellar models with $M_{\star}= 1.70 M_{\sun}$, 
$Z= 0.015$, $X^{\rm c}_{\rm H}= 0.3$ (central abundance)
and different assumptions for the value of $f$ of 
convective overshooting. The location of these models in the HR diagram are 
marked in Fig. \ref{figure-HROV} with black dots. The lower
panels of Fig. \ref{figure-rvxhOV} display the squared Brunt-V\"ais\"al\"a and Lamb frequency. As it can be seen, 
for larger values of $f$ (increasing overshooting), 
the H/He chemical interface becomes wider and less steep,
while shifts outward. The resulting feature in $N$, in turn, 
becomes wider and also shifts outward, following
the behavior of the $\mu$ gradient. 

\begin{figure}[h!] 
\begin{center}
\includegraphics[clip,width=9 cm]{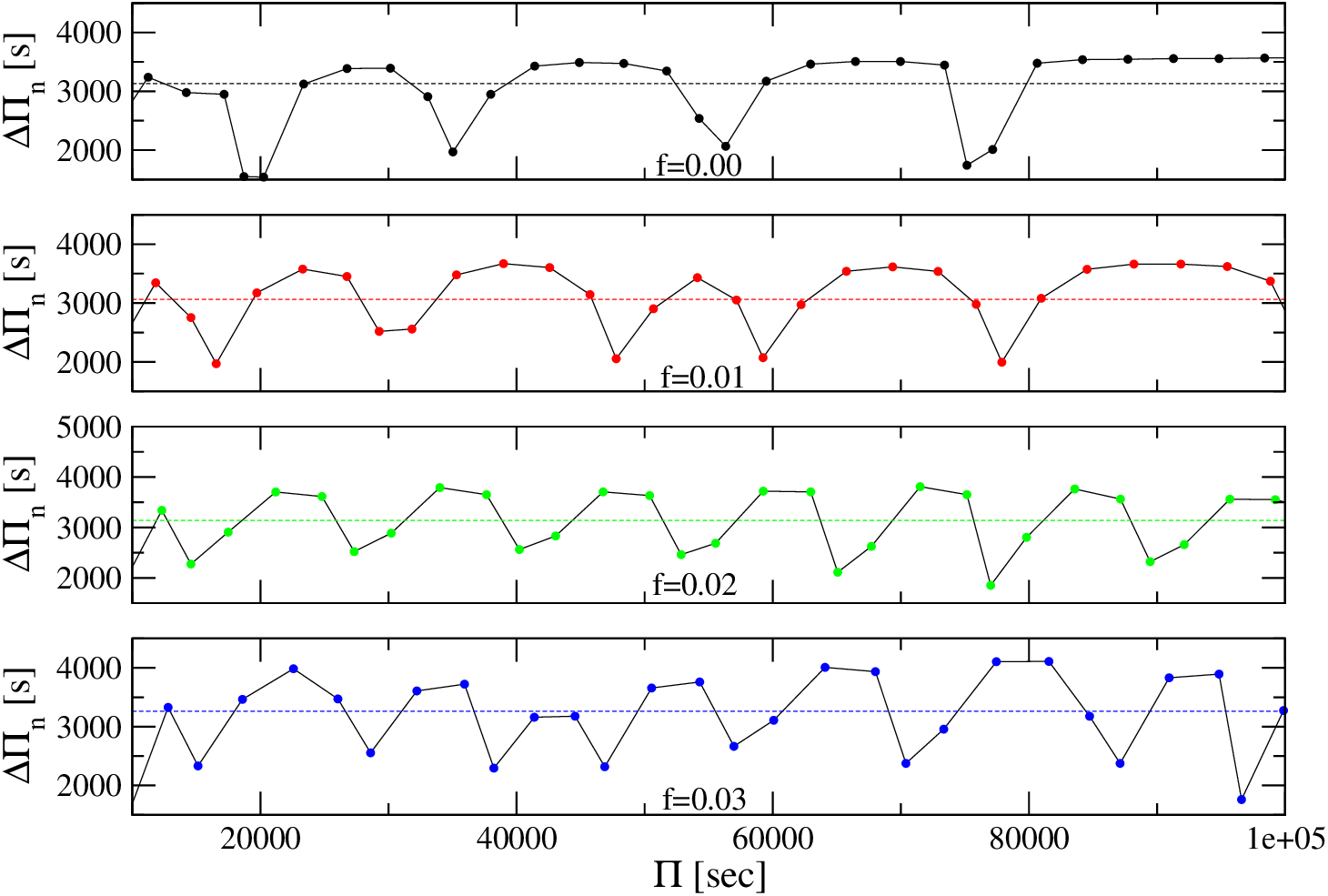} 
\caption{The forward period spacing 
($\Delta \Pi_n$) of dipole $g$ modes vs periods ($\Pi_n$) 
corresponding to the same stellar models shown in 
Fig. \ref{figure-rvxhOV}. The asymptotic period spacing 
($\Delta \Pi_n^{\rm a}$) is depicted with dashed lines in each case.}
\label{figure-DPvPOV} 
\end{center}
\end{figure} 

The impact of the different  assumptions for core overshooting in our
models on the forward period spacing of $g$-modes is displayed in
Fig. \ref{figure-DPvPOV}.  From upper to lower panels, the figure
shows the forward period spacing in terms of periods for dipole $g$
modes corresponding to  increasing overshooting ($f$ from 0 to
0.03). As it can be seen,  there are several minima of $\Delta \Pi_n$
present, whose number  increases with increasing the importance of
overshooting. The distinct behavior of $\Delta \Pi_n$  is due to that
not only the location but also the shape of the $\mu$ gradient are
strongly  modified when different values of $f$ are adopted (see
Fig. \ref{figure-rvxhOV}).  Also, we note that the asymptotic period
spacing slightly increases for larger $f$ values. This is because the
value of the integral  in Eq. (\ref{aps}) diminishes for larger $f$.
This behavior is in excellent agreement with the results  of
\citet{Miglio2008} (their Figure 17).

\begin{figure}[h!] 
\begin{center}
\includegraphics[clip,width=9 cm]{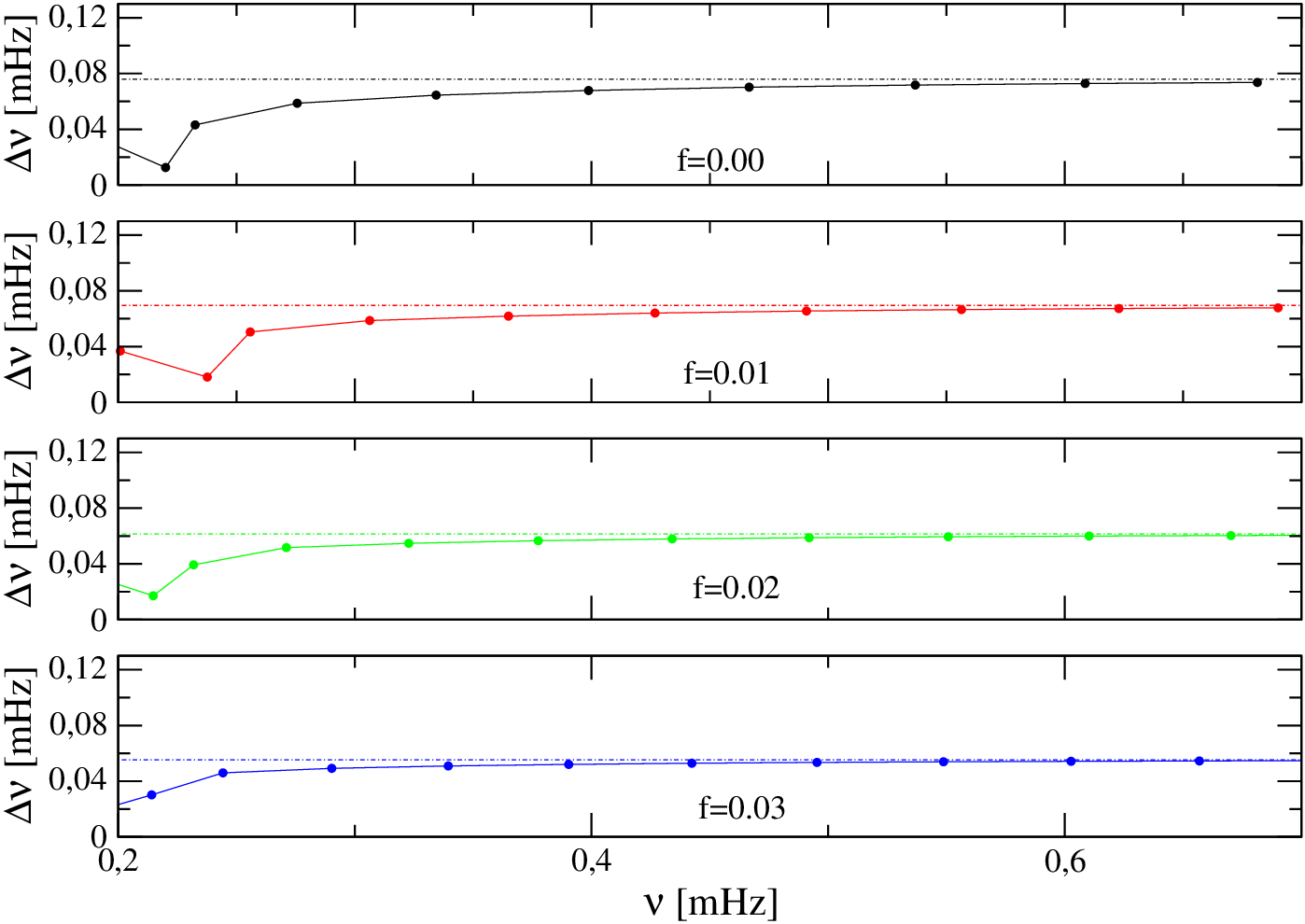} 
\caption{The forward frequency spacing 
($\Delta \nu$) of dipole $p$ modes vs frequency ($\nu$) 
corresponding to the same stellar models shown in 
Fig. \ref{figure-rvxhOV}. The asymptotic frequency spacing 
($\Delta \nu^{\rm a}$) is depicted with dashed lines in each case.}
\label{figure-DnvnOV} 
\end{center}
\end{figure}

Fig. \ref{figure-rvxhOV} shows the effect of different amount of
overshooting on the Lamb frequency. There exist a very tiny
discontinuity (small jump) in the frequency (not visible in the plot)
that moves with the edge of the convective core to outer regions, with
a growing overshooting parameter, and it extends along the chemical
composition gradient. This behavior does not significantly affects the
frequency spacing (see Fig. \ref{figure-DnvnOV}), whose structure
remains the same for each considered overshooting parameter. However,
in that figure, it can be observed that the asymptotic frequency
spacing, depicted with a horizontal line, slightly decreases as
larger overshooting parameters are considered. 

\begin{figure}[h!] 
\begin{center}
\includegraphics[clip,width=9 cm]{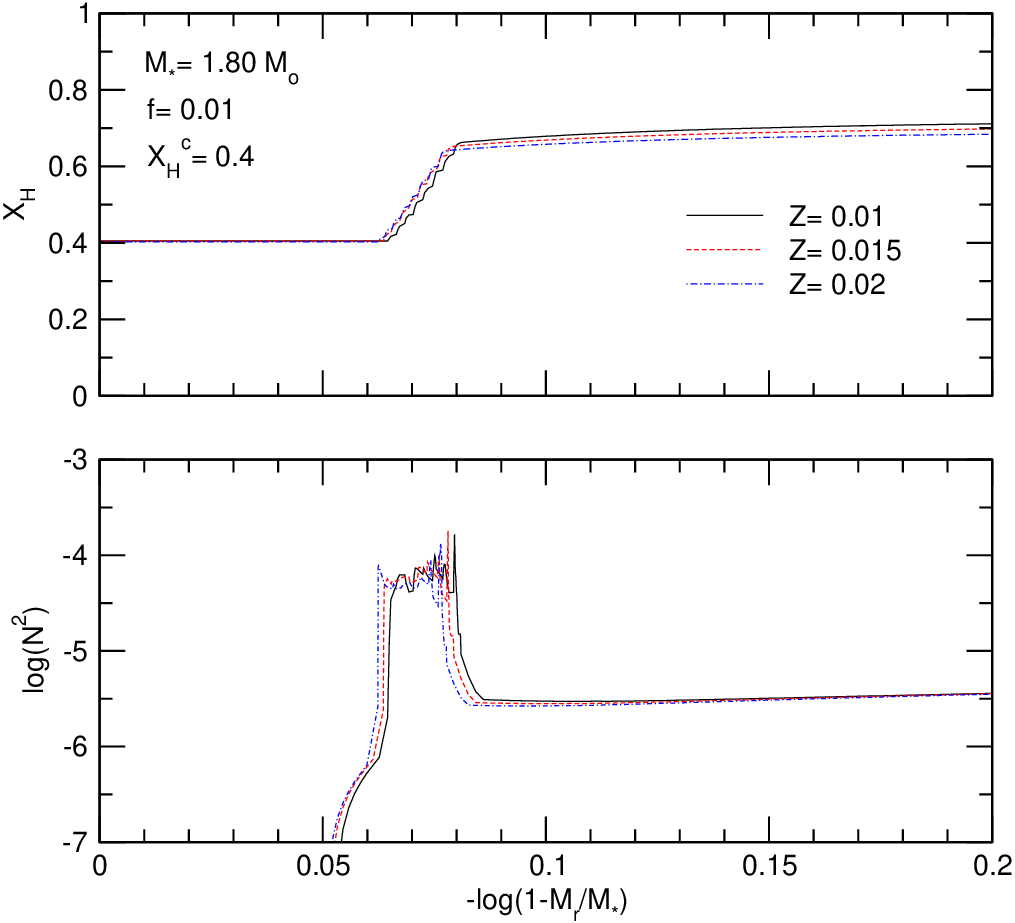} 
\caption{H abundance (upper panel), and the logarithm of the 
squared Brunt-V\"ais\"al\"a frequency (lower panel) corresponding 
to stellar models with $M_{\star}= 1.80 M_{\sun}$, 
$f= 0.01$, and $X^{\rm c}_{\rm H}= 0.4$ (central abundance), 
for different values of the metallicity, $Z= 0.01, 0.015, 0.02$. 
The location of these models in the HR diagram are 
marked in Fig. \ref{figure-HRZ} with black dots.}
\label{figure-rvxhz} 
\end{center}
\end{figure}

\begin{figure}[h!] 
\begin{center}
\includegraphics[clip,width=9 cm]{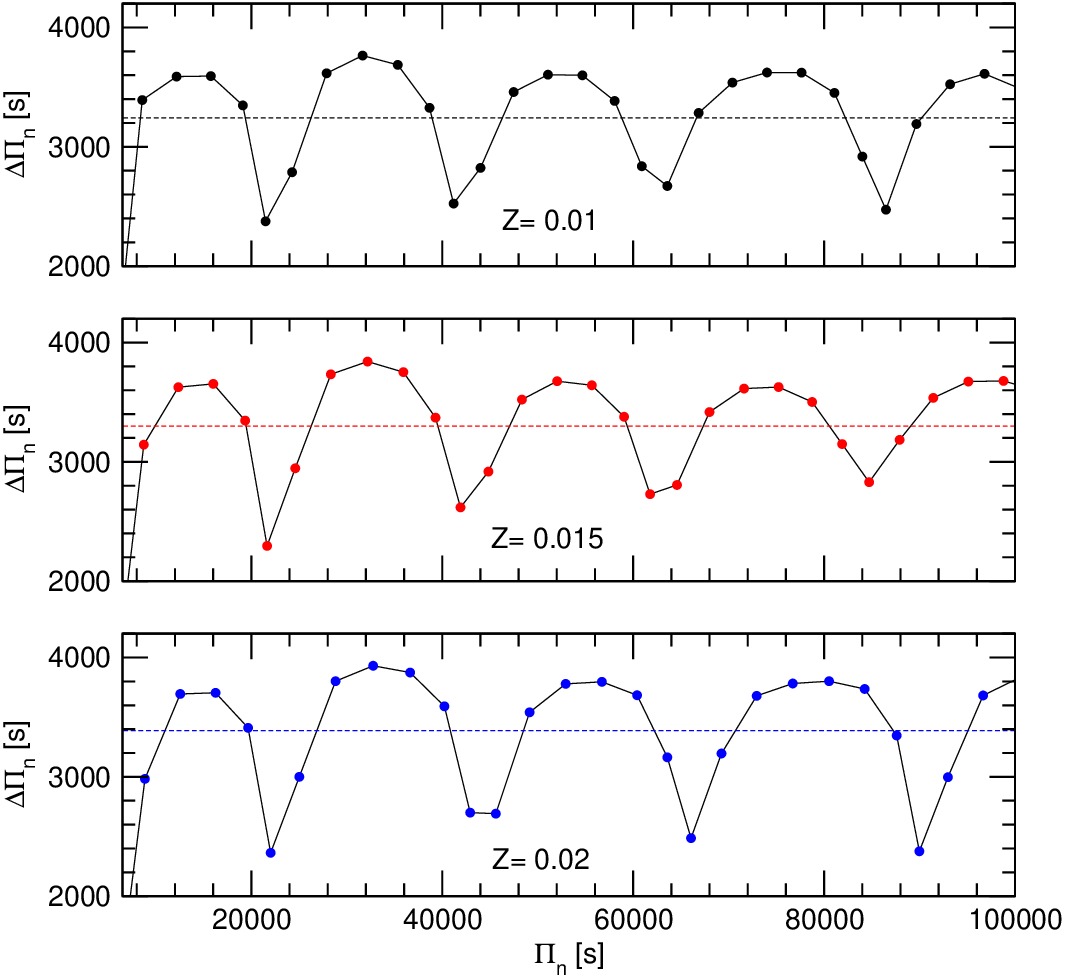} 
\caption{The dipole forward period spacing 
($\Delta \Pi_n$) of $g$ modes in terms of the periods 
($\Pi_n$) corresponding to the same stellar models 
shown in Fig. \ref{figure-rvxhz}. The asymptotic period spacing 
($\Delta \Pi_n^{\rm a}$) is depicted with dashed lines.}
\label{figure-DPvPz} 
\end{center}
\end{figure}

In closing this section, we briefly describe the effects of different
values for the metallicity $Z$ on the pulsation properties of our MS
models, while keeping fixed the stellar mass,  overshooting and the
central H abundance. Fig. \ref{figure-rvxhz} shows the H abundance and
the logarithm of the  squared Brunt-V\"ais\"al\"a frequency (upper and
lower panel, respectively)  for stellar models with $M_{\star}= 1.80
M_{\sun}$,  $f= 0.01$, and $X^{\rm c}_{\rm H}= 0.4$, and  different
values of the metallicity: $Z= 0.01, 0.015, 0.02$.  The location of
these models in the HR diagram is shown in  Fig. \ref{figure-HRZ} with
black dots. Note that different assumptions for the metallicity
---with the other parameters  fixed--- produces no appreciable change
in the chemical profile of  H and He, neither consequently in the
Brunt-V\"ais\"al\"a frequency.  As expected, there are no significant
differences in the behavior of the  period spacing ($\Delta \Pi_n$)
nor in the  the asymptotic period spacing ($\Delta \Pi_n^{\rm a}$) of
$g$ modes,  as shown in Fig. \ref{figure-DPvPz}. The same results (not
shown) are found for the frequency spacing and the asymptotic
frequency spacing of $p$ modes.

%::::::::::::::::::::::::::::::::::::::::::::::::

\section{Asteroseismological analysis}
\label{applications}

In this section we will describe our asteroseismological analysis on
the hybrid $\delta$ Sct-$\gamma$ Dor stars KIC 11145123
\citep{2014MNRAS.444..102K}, KIC 9244992 \citep{2015MNRAS.447.3264S},
HD 49434 \citep{2015MNRAS.447.2970B}, CoRoT 105733033
\citep{2012A&A...540A.117C}, and CoRoT 100866999
\citep{2013A&A...556A..87C}. As we shall see, we take fully advantage
of the information contained in both the acoustic- and the
gravity-mode period spectra offered by these target stars.

With the aim to searching for models that best reproduce the observed
pulsation spectra of each target star, we employed \textbf{two} different and
independent model-fitting procedures. We use the case of the HD 49434
star as an illustrative example  to describe them. The results along
with the characteristics of the other target stars are listed after
this description. 

HD 49434 has a visual magnitude $M_{\rm V}= 5.74$, with a 
FIV spectral type and effective temperature $T_{\rm eff}= 7632 \pm 126$ K
\citep{2004A&A...425..683B}. This hybrid $\delta$ Sct-$\gamma$ Dor
star is a rapid rotator, with velocities around $v \sin i= 85.4 \pm 6.6$
km s$^{-1}$ and a surface gravity of $\log g= 4.43 \pm 0.20$ estimated
by \citet{2006A&A...448..341G}. The radius of the star is 
$R_{\star}= 1.601 \pm 0.052 R_{\sun}$ according to 
\citet{2006A&A...450..735M} and its stellar mass is 
$M_{\star}= 1.55 \pm 0.14$ $M_{\sun}$ \citep{2002A&A...389..345B}. 
It is worth of mention that the hybrid status of this star is questioned in \citet{2009AIPC.1170..477B} and no gap between the frequency domains has been found in \citet{2012ASPC..462..111H}, probably due to its high rotation velocity, as it is mentioned in \citet{2015MNRAS.447.2970B}. Nevertheless, in \citet{2008A&A...489.1213U} and \citet{2011A&A...525A..23C} it is shown that HD 49434 has a
$\delta$ Sct period region with periods in the range $[1080-28800]$
sec, and also exhibit a $\gamma$ Dor period region with periods in the
range  $[28800-288000]$ sec. \citet{2015MNRAS.447.2970B} propose 
a main period spacing of $2030.4$ sec for $g$ modes. The largest 
amplitude period in the $p$-mode period range is at $9283.23$ sec. Table
\ref{tabladatosbrunsden} summarizes these stellar parameters and 
the observed pulsation period ranges for HD 49434.

\begin{table}[!ht]
  \centering
  \caption{Observational data for HD 49434.}
  \begin{tabular}{lc}
    \hline\hline\noalign{\smallskip}
          $T_{\rm eff}$ [K] &   $7632 \pm 126$ \\  
           $\log g$  &  $4.3 \pm 0.2$ \\
Period range of $\delta$ Sct domain [sec] & $[1080-28800]$ \\
Period range of $\gamma$ Dor domain [sec] & $[28800-288000]$ \\
  $\overline{\Delta \Pi}$ [sec] &  2030.4    \\ 
Period of the largest-amplitude mode [sec] &  9283.23 \\
$M_{\star} [M_{\rm \sun}]$ & $1.55 \pm 0.14$ \\
$R_{\star} [R_{\rm \sun}]$ & $1.601 \pm 0.052$\\
           \hline\hline
  \end{tabular}
  \label{tabladatosbrunsden}
\end{table}

\begin{itemize}

\item \underline{\textbf{Procedure 1}}: 
  
\textit{Step 1}: We calculated the mean
period spacing $\left(\overline{\Delta \Pi_n}\right)$ in the
observed $g$-mode period range for all the stellar models derived
from the numerical simulations. For this star we calculated
$\overline{\Delta \Pi_n}$ using $g$ modes with $\ell= 1$. In Figure
\ref{figureBrunsdenDP} we depict $\overline{\Delta\Pi_n}$
corresponding to the evolutionary phases from the ZAMS to the TAMS for
all the masses in our grid of models for the case with $Z= 0.01$ and
$f= 0.01$. The horizontal straight line is the observed mean period spacing 
of $g$ modes for this star $\left(\overline{\Delta \Pi}\right)$. As it can be
seen,  the mean period spacing decreases as the star evolves, thus
$\overline{\Delta\Pi}$ can be used as an indicator of the
evolutionary stage of stars and as a robust constraint in the
selection of models. In this step, we discard a large portion of the
grid of models ---those models that do not reproduce the observed
$g$-mode period spacing--- and we selected for each combination of 
$M_{\star}$, $Z$ and $f$ only the models that best reproduce the observed
$\overline{\Delta\Pi}$, keeping in total 252 models.  

\begin{figure}[h!] 
\begin{center}
\includegraphics[clip,width=9 cm]{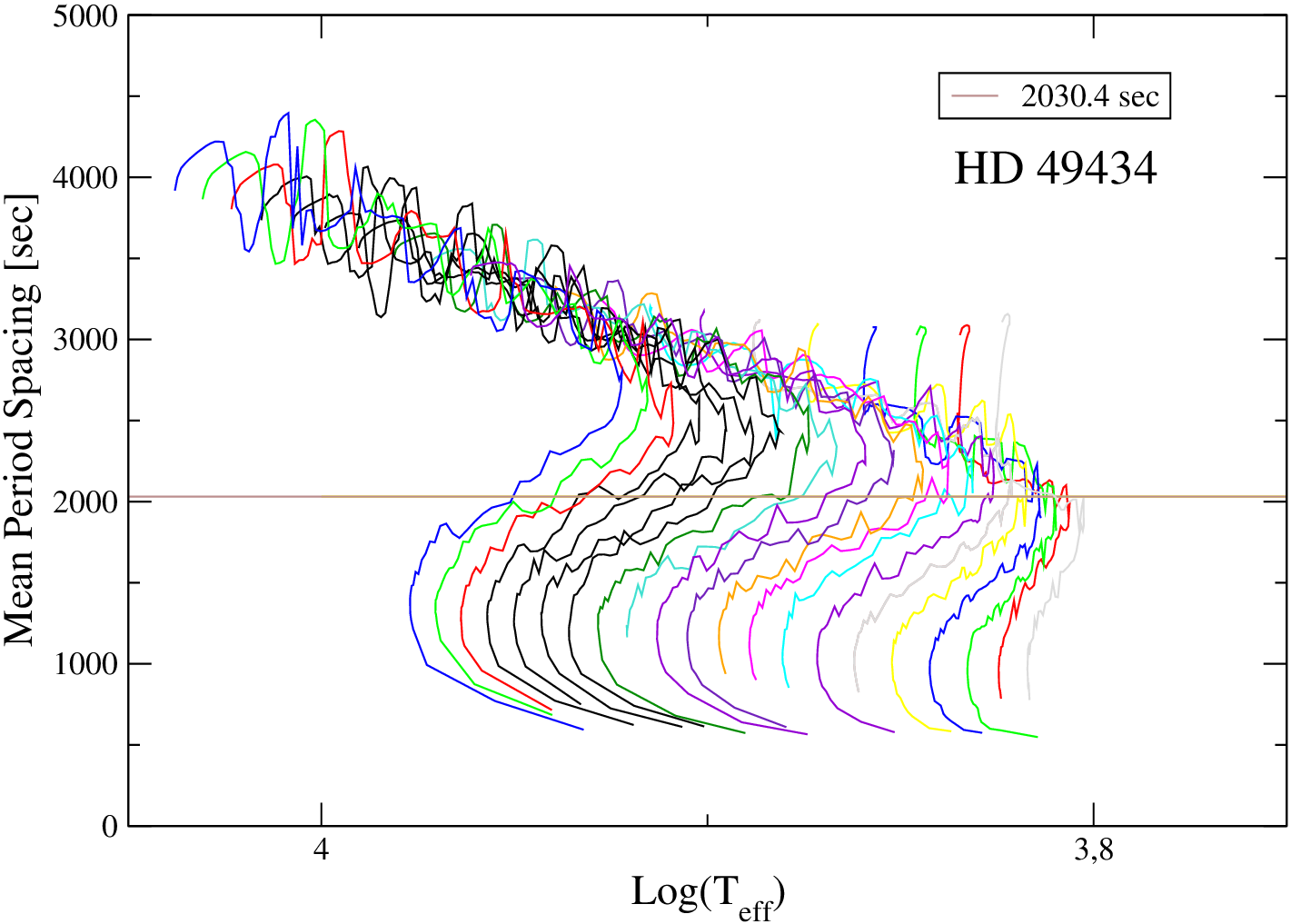} 
\caption{Mean period spacing vs. effective temperature for models with $Z=0.01$ and $f=0.01$. The straight line is the observed mean period spacing for $g$-modes corresponding to HD 49434.}
\label{figureBrunsdenDP} 
\end{center}
\end{figure} 

\textit{Step 2}: In this step, we assume that the largest amplitude
mode in the $\delta$ Sct region of the pulsation spectrum is
associated to the fundamental radial mode ($\ell= 0, n= 0$) or one
of the low radial overtone modes ($\ell= 0$, $n= 1, 2, 3, 4,\cdots$). This
allows us to reduce even more the number of possible seismological
models, by retaining only those models with the radial (fundamental or
an overtone) mode as close as possible to the observed largest
amplitude mode\footnote{In practice we adopted an arbitrary difference
  of $\sim 100$ s.}. In Fig \ref{figurebrunsdenradial} we depict the
periods associated to the fundamental radial mode ($n= 0$) and the
first radial overtone modes ($n= 1, 2, 3$) for models with $Z= 0.01$ and
$f= 0.01$ selected in the previous step, e. g. those models that best
reproduce $\overline{\Delta\Pi}$. The horizontal straight line
indicates the period associated to the largest-amplitude mode in the
$\delta$ Sct region of the pulsation spectrum of  HD 49434
(i.e. $9283.23$ sec). 

\begin{figure}[h!] 
\begin{center}
\includegraphics[clip,width=9 cm]{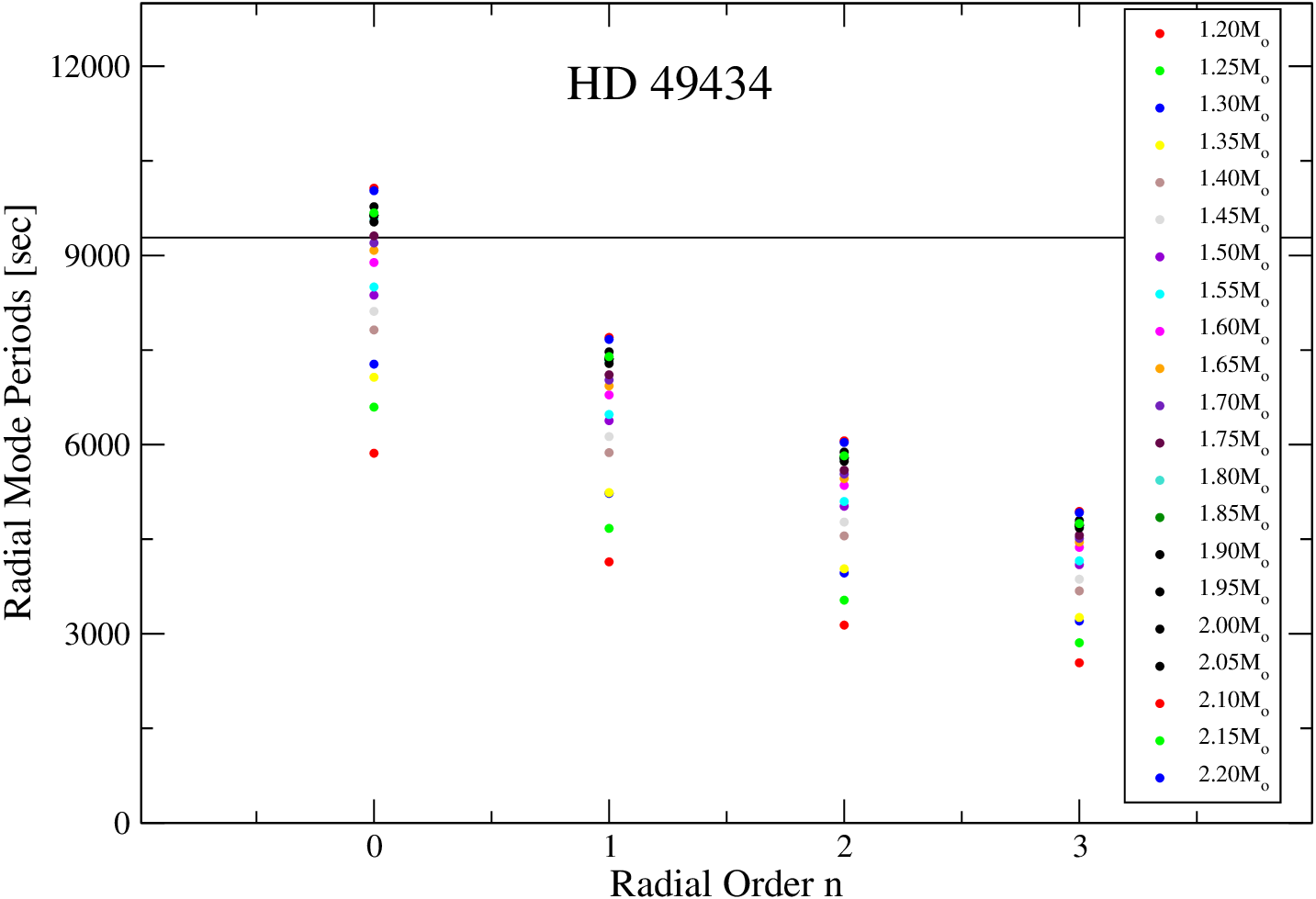} 
\caption{Radial mode periods vs. radial order $n$ for models selected in the previous step with $Z=0.01$ and $f=0.01$ for HD 49434.}
\label{figurebrunsdenradial} 
\end{center}
\end{figure} 

\textit{Step 3}: For the models selected in the previous steps, we
performed a period-to-period fit to the observed $p$-mode periods.  We
calculated the quantity:

\begin{equation}
\chi^p=\sum_{i=1}^{n}{ \frac{\left[\Pi_{\rm o}^p-\Pi_{\rm c}^p\right]_i^2}
{\left( \sigma_{\Pi^p}^2 \right)_i}},
\end{equation}

\noindent where  $\Pi^p_{\rm o}$ is the observed period, $\Pi^p_{\rm c}$ is the
calculated period, $\sigma_{\Pi^p}$ is the uncertainty associated with
the observations of the considered $p$-mode period, and $n$ is the
total number of observed $p$-mode periods. Since no identification of the 
harmonic degree is available at the outset for HD 49434, we
calculated $\chi^p$ by searching for the best
period fit among periods of $p$ modes with $\ell= 1, 2$ and
$3$. Accordingly, for HD 49434, $\Pi^p_{\rm o}$ takes its value among 
the periods classified as DSL (``\textbf{$\delta$} Scuti like'')
in the $\delta$ Sct domain in
Table 3 of \citet{2015MNRAS.447.2970B} and $\Pi^p_{\rm c}$ is the calculated
period for $\Pi^p_{\rm o}$, which is a non-radial $p$ mode with harmonic
degree $\ell= 1, 2$ or $3$. All the uncertainties in the period observations
were calculated considering the conservative error of $0.01$ d$^{-1}$
mentioned in \citet{2015MNRAS.447.2970B}.

\textit{Step 4}: Finally, in order to obtain the ``best-fit model''
for each target star, we calculated the quantity $F_1$ for all
the selected models in \textit{Step 2}:

\begin{equation}
F_1=\sum_{j=1}^{N}{\frac{\left[\overline{\Delta\Pi}_n-
\overline{\Delta\Pi} \right]_j^2}
{\left(\sigma_{\overline{\Delta\Pi}}^2\right)_j}+
\frac{\left[\Pi_{\rm o}^r-\Pi_{\rm c}^r\right]_j^2}{(\sigma_{\Pi^r}^2)_j}+
\chi_j^p}
\label{f1}
\end{equation}

\noindent This quantity takes into account the differences between the
calculated and observed mean period spacing ($\overline{\Delta\Pi}_n$
and $\overline{\Delta\Pi}$ respectively), the difference between the
calculated radial (fundamental or an overtone) modes $\Pi_{\rm c}^r$ and
the observed largest amplitude mode $\Pi_{\rm o}^r$, and the quantity
$\chi_j^p$ calculated in the previous step, which relates the observed
set of $p$-mode periods with the calculated one. In this expression,
$N$ is the number of selected models in the \textit{Step 2}, e. g.,
those ones which nearly reproduce the observed $\overline{\Delta\Pi}$
and the largest amplitude mode. $\sigma_{\overline{\Delta\Pi}}$ is the
observational uncertainty for  $\overline{\Delta\Pi}$, and
$\sigma_{\Pi^r}$ is the uncertainty corresponding to the largest-amplitude
mode in the target star. 

\item \underline{\textbf{Procedure 2}}:

In this procedure we considered the possibility of having radial modes
among the $p$-mode range for the target stars. With this assumption,
we exclude the direct association of the largest amplitude mode in the
$p$-mode range with a radial mode, which is usually done for HADS
stars \citep{2015pust.book.....C}. We performed a comparison between
the observed periods in the $p$-mode range and the theoretical radial
and nonradial $p$-mode periods obtained from the numerical
models. 

The \textit{Step 1} is the same as in \textbf{Procedure 1}, i. e. we selected all models that best reproduce the
observed $\overline{\Delta \Pi}$ of $g$ modes.

\textit{Step 2}: We calculated the quantity $\chi^{rp}$, for
models selected in the previous steps, in order to compare the
observed periods in range of $p$-mode periods of the target star with
the periods of radial and nonradial $p$-modes: 

\begin{equation}
\chi^{rp}=\sum_{i=1}^{n}{ \frac{\left[\Pi_{\rm o}^{rp}-\Pi_{\rm c}^{rp}\right]_i^2}{(\sigma_{\Pi^{rp}}^2)_i}},
\label{chirp}
\end{equation}

\noindent where $\Pi_{\rm o}^{rp}$ stand for the observed periods in the
range of $p$ modes for the target star, $\Pi_{\rm c}^{rp}$ is the calculated
period that best fits $\Pi_{\rm o}^{rp}$, which could be a radial or
nonradial $p$ mode, $\sigma_{\Pi^{rg}}$ is the uncertainty associated
with the observations of the considered $p$-mode or radial-mode period, and
$n$ is the total number of observed $p$-mode periods for the target star.

\textit{Step 3}: We calculated the following quantity for the models selected in the \textit{Step 1}:

\begin{equation}
\chi^g=\sum_{i=1}^{n}{\frac{\left[\Pi_{\rm o}^g-\Pi_{\rm c}^g\right]_i^2}{(\sigma_{\Pi^g}^2)_i}}
\label{chig}
\end{equation}

\noindent where $\Pi^g_{\rm o}$ is the observed period, $\Pi^g_{\rm c}$ is the calculated period, $\sigma_{\Pi^g}$ is the uncertainty associated with 
the observations of the considered $g$-mode periods and $n$ is the total number of observed $g$-mode periods. As it was done for calculating $\chi^p$,  we search for the best period fit between the observed periods of $g$-modes and theoretical periods with $\ell= 1, 2, 3$. Thus, for HD 49434, $\Pi^g_{\rm o}$ takes its value from the periods in the $\gamma$ Dor domain, classified as GDL ($\gamma$-Doradus like”) in Table 3 of \citet{2015MNRAS.447.2970B}.

\textit{Step 4}: Finally, we selected the ``best-fit model'' by
calculating the quantity $F_2$ between all models previously
selected, incorporating $\chi^{rp}$ and $\chi^g$:

\begin{equation}
F_2=\sum_{j=1}^{N}{\frac{\left[\overline{\Delta\Pi}_n-\overline{\Delta\Pi} \right]_j^2}{\left(\sigma_{\overline{\Delta\Pi}}^2\right)_j}+\chi^{rp}_j+ \chi^g_j}
\end{equation}

\noindent It is worth mentioning that, for HD 49434, we did not
include in $\chi^p$ the period classified as DLS with the largest
amplitude. Instead, we assumed that it is a period corresponding to a
radial mode in order to perform the \textit{Step 2} of
\textbf{Procedures 1}.  In \textbf{Procedure 2} we
included in Eq.(\ref{chirp}) the aforementioned largest-amplitude
mode, i.e, we considered the possibility of having radial modes in the
$\delta$ Sct domain. Table \ref{modelosbrunsdenprocedures} resumes the
characteristics of the models selected for each procedure.

\begin{table}[!ht]
  \centering
  \caption{Best-fit models for HD 49434.}
  \begin{tabular}{lcc}
    \hline\hline\noalign{\smallskip}
                 & \textbf{Procedures 1} & \textbf{Procedure 2} \\ 
\hline\noalign{\smallskip}

    $M_{\star} [M_{\rm \sun}]$ &  1.75   &  1.75  \\ 
    $Z$                      &  0.01   &  0.015 \\
    $f$                      &  0.01   &  0.03  \\
    $T_{\rm eff}$ [K]         &  7399   &  6504 \\
    $\log g$                 &  3.85   &  3.51 \\
    $R_{\star} [R_{\rm \sun}]$ &  2.57   & 3.81  \\
    Age[$10^6$ yr]           &  1169.08 &  1641.97\\
    $L_{\star}[L_{\rm \sun}]$  &  19.36   &  24.33\\
    $\overline{\Delta \Pi}$ [sec]&   2045.42 & 2026.92 \\
               \hline
 \end{tabular}
  \label{modelosbrunsdenprocedures}
\end{table}
\end{itemize}

Summarizing we obtained one model with $Z= 0.01$ and $f= 0.01$ from
\textbf{Procedure 1}, and another  model with $Z= 0.015$
and $f= 0.03$ from \textbf{Procedure 2}. We perform \textbf{Procedure 2}
 for this star since the mode classification is not conclusive and
therefore the existence of radial modes in the $\delta$ Sct region is
possible. Both models have the same mass: $1.75 M_{\sun}$, the
one with $f= 0.03$ is reaching the TAMS, and the other one with $f= 0.01$ is
before the ``knee'' of the MS, as can be observed in
Fig. \ref{figurerhmodelosfinales}. Besides, we incorporated a study of the observed individual $g$-mode periods of the target star in the Eq. \ref{f1}, by adding the $\chi^g$ term described in Eq. \ref{chig}. The ``best fit model'' remained the same without this term, thus adding
a period-to-period fit for the $g$-modes does not affect this selection. Also, when we include the
possibility of having radial modes in the $\delta$ Sct period domain,
we obtain another model. This reinforces the need of having a correct
mode classification. 

Next we will detail these procedures for the other four target stars.
   
\subsection{KIC 11145123} %Kurtz

KIC 11145123, also known as the ``Holy Grail'', is a late A $\delta$
Sct-$\gamma$ Dor hybrid star observed by the \emph{Kepler} Mission and
extensively studied by \citet{2014MNRAS.444..102K}. From the
\emph{Kepler} Input Catalogue (KIC) revised photometry
\citep{2014ApJS..211....2H}, its effective temperature is $8050 \pm 200$ K
and its surface gravity is $\log g= 4.0 \pm 0.2$ (cgs units). It
also has a \textit{Kepler} magnitude $Kp= 13$. The $p$-mode periods
present in  KIC 11145123 are in the range $[3536.75-5160.67]$ sec and
the $g$-mode period range is $[42255.53-71430]$ sec. The complete list
of $p$- and $g$-mode frequencies are in Table 2 and 3 respectively of
\citet[][]{2013A&A...556A..52T}. The mean period spacing for $g$ modes
is $\overline{\Delta \Pi}= 2073.6$ sec and the largest amplitude mode
in the $\delta$ Sct region of the pulsation spectrum has a period of
4810.69 sec. Table \ref{tabladatoskurtz} resumes these stellar
parameters and the
observed pulsation-period range for KIC 11145123.

\begin{table}[!ht]
  \centering
  \caption{Observational data for KIC 11145123.}
  \begin{tabular}{lc}
    \hline\hline\noalign{\smallskip}
          $T_{\rm eff}$ [K] &   8050 \\
            $\log g$      &  $4.0 \pm 0.2$ \\
Period range of $\delta$ Sct domain [sec] &  $[3536.75-5160.67]$ \\
Period range of $\gamma$ Dor domain [sec] & $[42255.53-71430]$ \\ 
       $\overline{\Delta \Pi}$ [sec]& 2073.6   \\
Period of the largest amplitude mode [sec]& 4810.69 sec   \\
           \hline\hline 
  \end{tabular}
  \label{tabladatoskurtz}
\end{table}

\begin{figure}[h!] 
\begin{center}
\includegraphics[clip,width=9 cm]{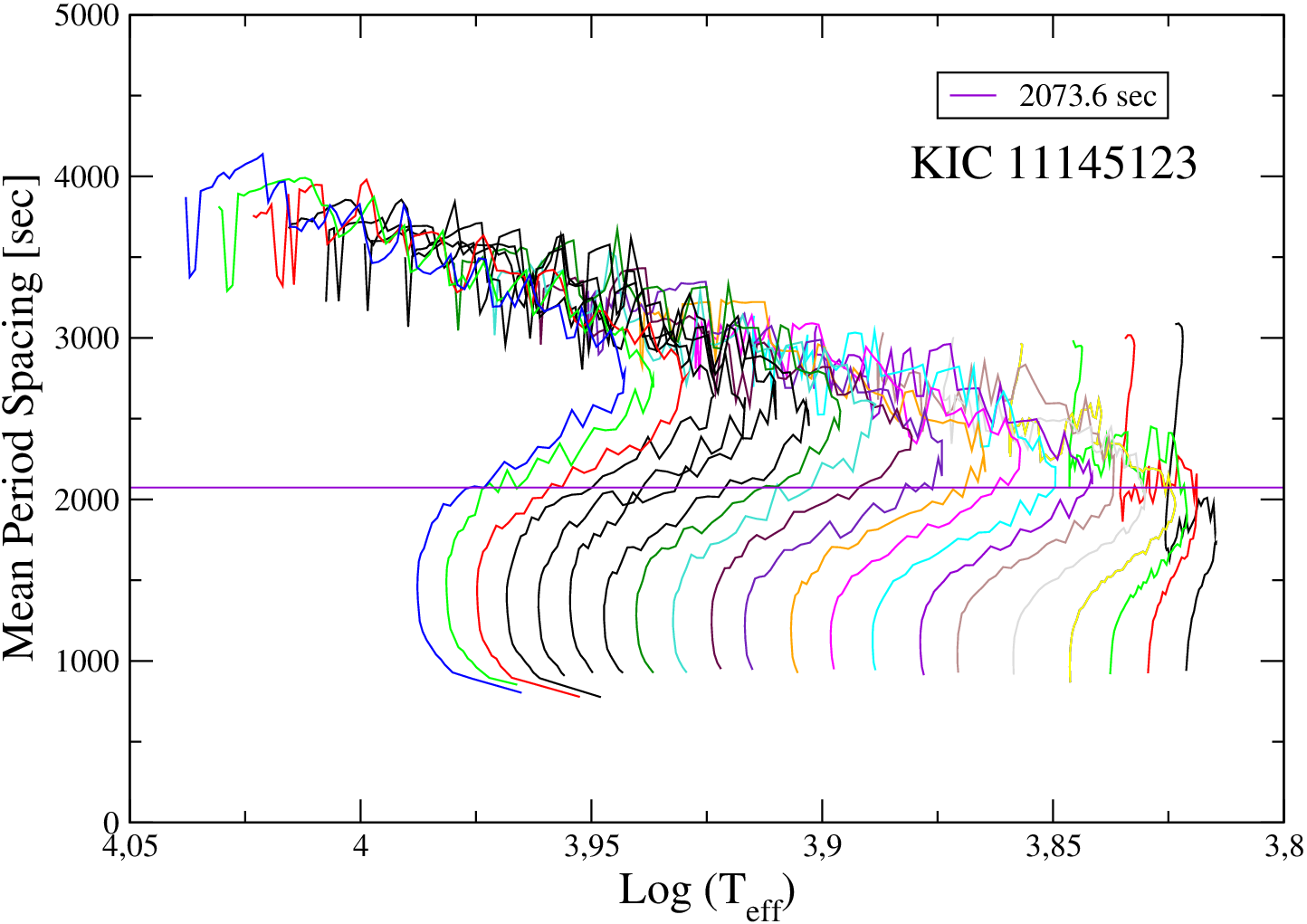} 
\caption{Mean period spacing vs. effective temperature for
  models with $Z= 0.01$ and $f= 0$. The straight line is the observed
  mean period spacing for $g$-modes for the case of KIC 11145123.}
\label{figureKurtzDP} 
\end{center}
\end{figure}

In the \textbf{Procedure 1}, for the \textit{Step 1}, we calculated
$\overline{\Delta \Pi_n}$ in the $\gamma$ Dor domain using only
pulsation $g$ modes with harmonic degree $\ell= 1$, since KIC 11145123
shows only triplets in the $\gamma$ Dor domain. Figure
\ref{figureKurtzDP} shows the mean period spacing for models with
$Z= 0.01$ and $f= 0$.

\begin{figure}[h!] 
\begin{center}
\includegraphics[clip,width=9 cm]{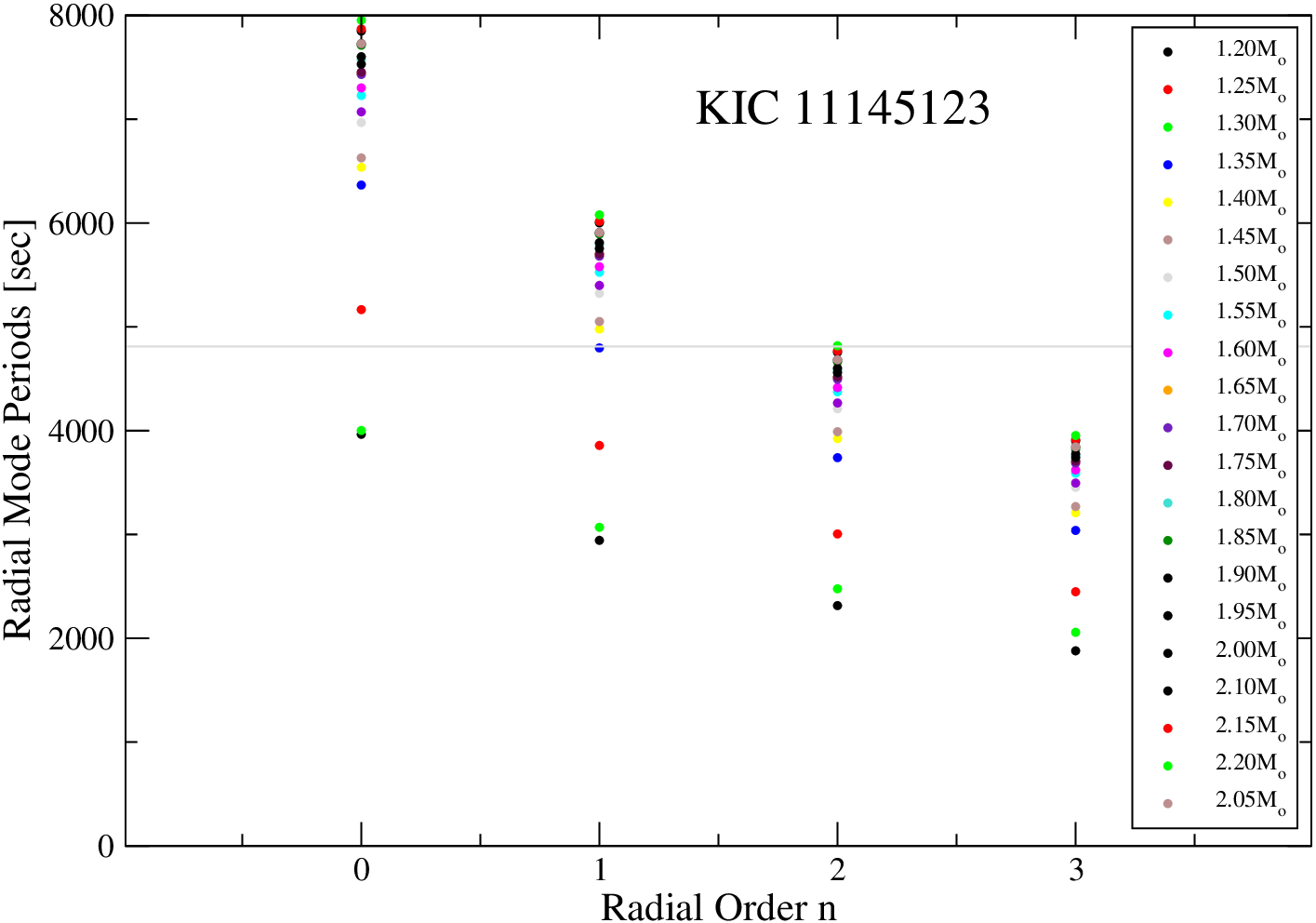} 
\caption{Radial mode periods vs radial order $n$ for models selected
  in the \textit{Step 1}  with $Z=0.01$ and $f=0$ corresponding
  to KIC 11145123.}
\label{figureKurtzradial} 
\end{center}
\end{figure}

We performed \textit{Step 2} of \textbf{Procedure 1}. We depict in
Figure \ref{figureKurtzradial} the radial mode periods (the fundamental
and the overtones) for models with $Z= 0.01$ and $f= 0$
selected from the previous step. 

We modified \textit{Step 3} ---given that this target star has modes
classified with $\ell= 1$ and $\ell= 2$--- computing two quantities:
$\chi_1$ and $\chi_2$, for models with $p$-mode periods with $\ell= 1$
and $\ell= 2$ respectively, since KIC 11145123 exhibits three
rotational quintuplets ($\ell= 2$) and five rotational
triplets ($\ell= 1$). The expression for these quantities is given by:

\begin{equation}
\chi_j= \sum_{i=1}^{N}\frac{\left(\Pi_{\rm o}^{p,j}-\Pi_{\rm c}^{p,j} \right)_i^2}{\left(\sigma_{\Pi}^2\right)_i},
\end{equation}

\noindent where $N$ is the total number of observed $p$-mode periods
with $\ell= 1$ for $j= 1$ and $\ell= 2$ for $j= 2$. $\Pi_{\rm o}^{p,j}$ is
the observed $p$-mode period, $\Pi_{\rm c}^{p,j}$ is the closest period to
$\Pi_{\rm o}^{p,j}$ between those ones with $\ell= 1$ for $j= 1$ and
$\ell= 2$ for $j= 2$ and $\sigma_{\Pi}$ is the uncertainty associated with
$\Pi_{\rm o}^p$. Thus, $\chi^p$ is given by $\chi^p=\chi_1+\chi_2$.

As it has been done for HD 49434, we included a period-to-period fit for the $g$-modes. We used the $g$-mode periods listed in Table 3 of \citet{2014MNRAS.444..102K} in order to calculate $\chi^g$ and we obtained two different models.

\textbf{Procedure 2} was not performed
for this star since all its frequencies in $\delta$ Sct region were
classified as $p$-mode with their corresponding harmonic degree and we
believe this classification is correct.

\begin{table}[!ht]
  \centering
  \caption{Best-fit models for KIC 11145123 with \textbf{Procedure 1}}.
  \begin{tabular}{lcc}
    \hline\hline\noalign{\smallskip}
                 & \textbf{Without $\chi^g$} & \textbf{With $\chi^g$}  \\ \hline\noalign{\smallskip}
    $M_{\star} [M_{\rm \sun}]$  &   1.35      &   2.2  \\ 
    $Z$                       &    0.01      &   0.01  \\
    $f$                       &    0.03      &   0      \\
    $T_{\rm eff}$ [K]          &    6064      &  9402         \\
    $\log g$                  &    3.77    & 3.98       \\
    $R_{\star} [R_{\rm \sun}]$  &    2.50    &  2.50 \\
    Age[$10^6$ yr]            &    3219.02   & 549.61      \\
    $L_{\star} [L_{\rm \sun}]$  &    7.75  & 48  \\
    $\overline{\Delta \Pi}$ [sec] &     2085.17  &  2071.87  \\
            \hline
  \end{tabular}
  \label{modeloskurtzprocedures}
\end{table}

In Table \ref{modeloskurtzprocedures}, the characteristics of the
asteroseismological models selected for KIC 11145123 according
\textbf{Procedure 1} with and without a $g$-mode period-to-period fit are shown. Two different seismological models were obtained meaning that individual $g$-mode period fits play an important role in the modeling of this
star.  Comparing our models with those obtained in
\citet{2014MNRAS.444..102K} with  $M_{\star}= 1.40 M_{\sun},
1.46 M_{\sun}$ and $2.05 M_{\sun}$ and $Z= 0.01$ and
$0.014$, we note some differences in these values, possibly due to
the fact that we did not considered atomic diffusion in our
simulations (the Brunt-V\"ais\"al\"a frequency is modified by this
physical process). Nevertheless, our best-fit models are both in the
TAMS overall contraction phase in good agreement with those obtained in
\citet{2014MNRAS.444..102K} and also have a metallicity  consistent
with the hypothesis that this star could be a SX Phe star. Note that
none of both models reproduce well the reported effective temperature
for this star. 

\subsection{KIC 9244992} %Saio

This star, observed by the \emph{Kepler} Mission, has an effective
temperature of $T_{\rm eff}= 6900 \pm 300$ K and its surface gravity is
$\log g=3.5 \pm 0.4$ according to the KIC revised
photometry \citep{2014ApJS..211....2H}. A recent study
\citep{2015EPJWC.10106049N} obtains $3.8388 \le \log T_{\rm eff} \le
3.8633$ K and $3.5 \le \log g \le 4.0$ and a rotational velocity of
$v\sin(i) \le 6$ km sec$^{-1}$ which indicates KIC 9244992 is a slow
rotator. The $p$- and $g$-mode period ranges for this star are
$[4689.89-7017.28]$ sec and $[54000-96000]$ sec, respectively. It shows a
radial mode with a period equal to 7001.97 sec and its mean period
spacing for $g$-modes is $\overline{\Delta \Pi}= 2280.96$ sec. The
complete list of $g$- and $p$-mode periods are shown in Table 1 and 3,
respectively, of \citet{2015MNRAS.447.3264S}. Table \ref{tabladatossaio}
summarizes these stellar parameters and the observed
pulsational periods range for KIC 9244992.

\begin{table}[!ht]
  \centering
  \caption{Observational data for KIC 9244992.}
  \begin{tabular}{lc}
    \hline\hline\noalign{\smallskip}
          $T_{\rm eff}$ [K] &  $6900 \pm 300$ \\
            $\log g$  & $3.5 \pm 0.4$  \\
Period range of $\delta$ Sct domain [sec] &  $[4689.89-7017.28]$ \\
Period range of $\gamma$ Dor domain [sec] &  $[54000-96000]$ \\ 
  $\overline{\Delta \Pi}$ [sec] &   2280     \\
Period of the largest-amplitude mode [sec]& 7002.18 \\
           \hline\hline
  \end{tabular}
  \label{tabladatossaio}
\end{table}

\begin{figure}[h!] 
\begin{center}
\includegraphics[clip,width=9 cm]{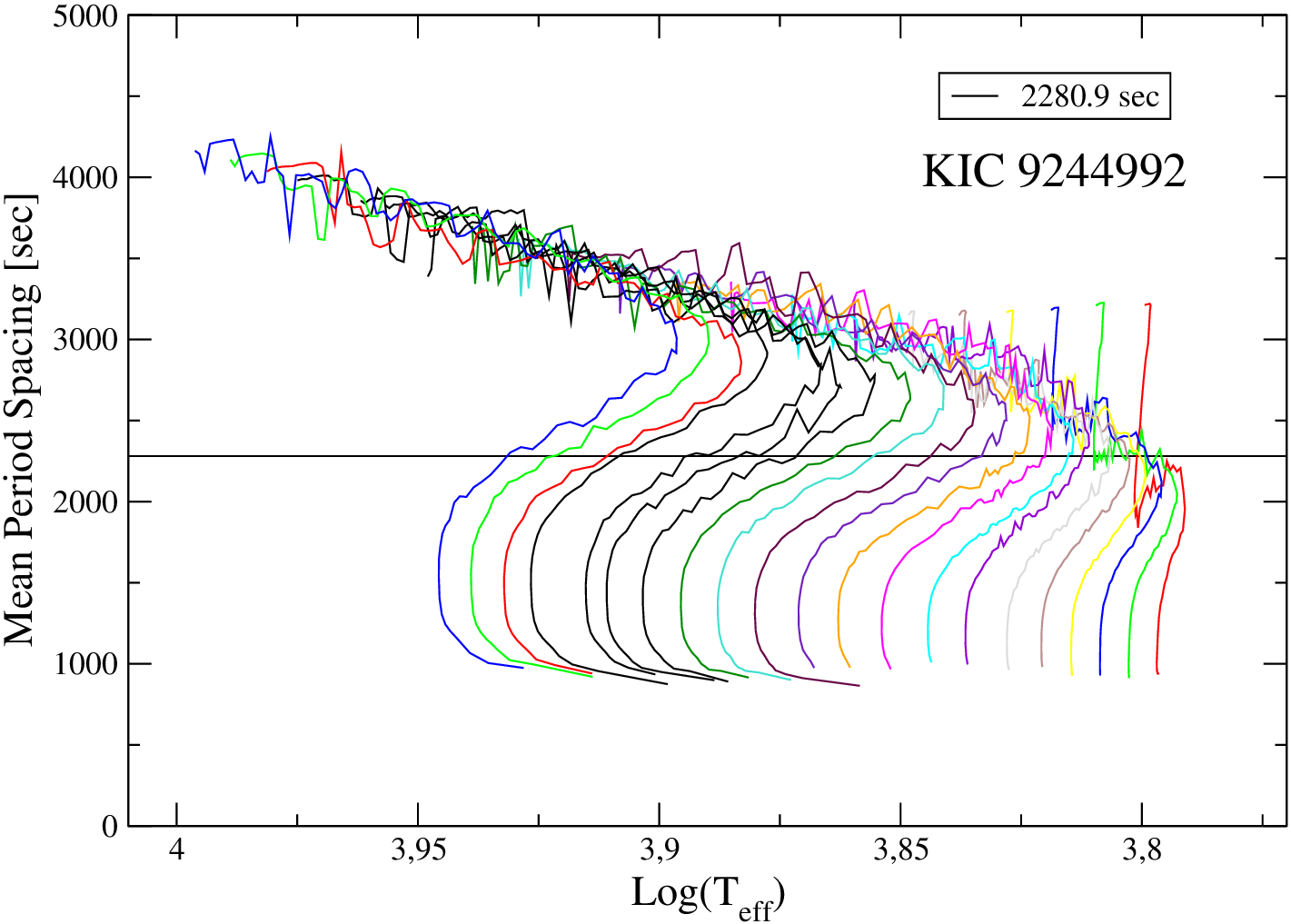} 
\caption{Mean period spacing of $g$ modes vs effective temperature for models
  with $Z= 0.02$ and $f= 0$. The straight horizontal line is the observed mean
  period spacing for $g$-modes corresponding to KIC 9244992.}
\label{figureSaioDP} 
\end{center}
\end{figure}

We calculated the mean period spacing in the observed $g$-mode period
range ($\overline{\Delta \Pi_n}$) using models with harmonic degree
$\ell= 1$. In Figure \ref{figureSaioDP} we depict the calculated
$\overline{\Delta \Pi_n}$ vs. $\log T_{\rm eff}$ for models with
$Z= 0.02$ and $f= 0$ and different values of stellar mass.

\begin{figure}[h!] 
\begin{center}
\includegraphics[clip,width=9 cm]{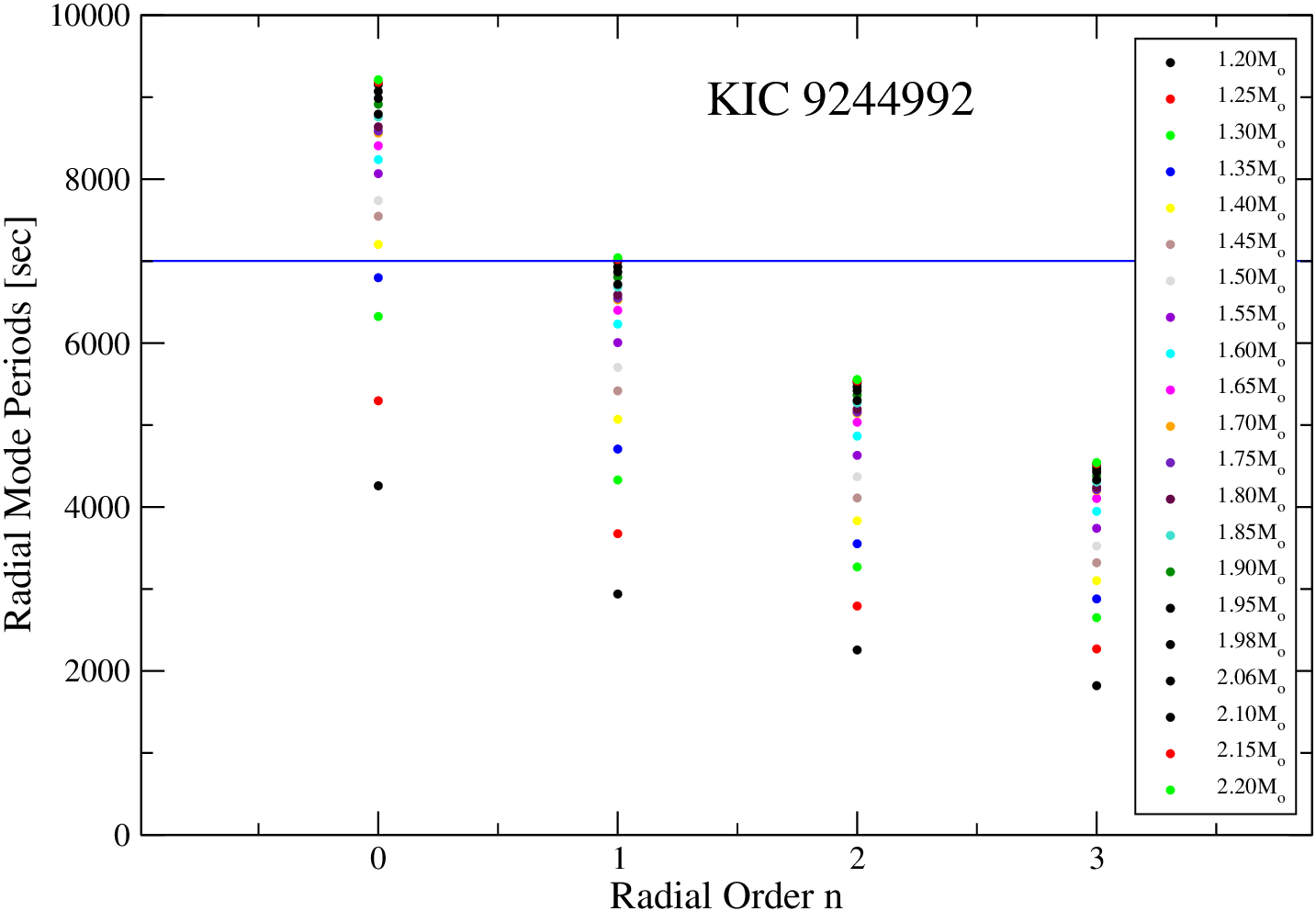} 
\caption{Radial mode periods vs radial order $n$ for the models
  selected in the previous step with $Z= 0.02$ and $f= 0$, corresponding
  to KIC 9244992. }
\label{figuresaioradial} 
\end{center}
\end{figure}

Next, we perform the \textit{Step 2} of \textbf{Procedure 1}. Figure
\ref{figuresaioradial} shows the radial mode periods for the selected
models in the previous step with $Z= 0.02$ and $f= 0$. 

KIC 9244992 has six $p$-mode frequencies with $\ell= 1$ and four
without assigned harmonic degree, according to the classification in
\citet{2015MNRAS.447.3264S}. Thus, we modified \textit{Step 3} with
the aim of include this classification. We calculated two quantities.
On one hand, we asess $\chi_1$, which compares
$\nu_2$, $\nu_7$, $\nu_8$, $\nu_9$,
$\nu_{11}$ and $\nu_6$ ($\Pi^p_{\rm o}$) from
Table 3 of \citet{2015MNRAS.447.3264S} with the calculated
$p$-mode periods with harmonic degree $\ell= 1$, by means of the
following expression:

\begin{equation}
\chi_{1}=\sum_{i=1}^{6}{ \frac{\left[\Pi^p_{\rm o}-\Pi^p_{\rm c}\right]_i^2}{(\sigma_{1}^2)_i}},
\label{chi1}
\end{equation}

\noindent where $\sigma_1$ is the uncertainty associated with the
observed  $p$-mode frequencies with $\ell= 1$. On the other hand,
we compute $\chi_{123}$, which compares the frequencies without
an assigned harmonic degree ($\Pi^p_{\rm o}$) with those  $p$-mode
frequencies calculated with $\ell=1, 2$ and $3$ ($\Pi^p_{\rm c}$), that is,
we search for the best period-fit  among the non-radial $p$ modes with
$\ell= 1, 2$ and $3$. Thus, $\chi_{123}$ is defined as:

\begin{equation}
\chi_{123}=\sum_{i=1}^{4}{ \frac{\left[\Pi^p_{\rm o}-\Pi^p_{\rm c}\right]_i^2}{\left(\sigma_{123}^2\right)_i}}
\label{chi123}
\end{equation}

\noindent where $\sigma_{123}$ is the uncertainty associated with the
observations of the considered $\Pi^p_{\rm o}$. Thus, $\chi_p$ has this
expression: $\chi_p=\chi_{1}+\chi_{123}$.

\textbf{Procedure 2} was not performed since
all the $p$-mode periods are classified with their corresponding harmonic
degree.

In Table \ref{modelossaioprocedures} we show the characteristics of
the best-fit model selected for KIC 9244992. Also in this case, we added a $g$-mode period-to-period fit in Eq.\ref{f1}, considering  models with $\ell= 1$ in order obtain $\chi_g$ since the target star has only rotational triplets in the $g$-mode period range. The same model was obtained, meaning that including a $g$-mode period-to-period fit did not affect the selection of the model. In
\citet{2015MNRAS.447.3264S} the authors propose a model with
$M_{\star}= 1.45M_{\sun}$, $Z= 0.01$ and $f= 0.005$. Our $2.1M_{\sun}$
best-fit model is more massive, does not have overshooting in the core
and has higher metallicity ($Z= 0.02$). Thus, our results do not
indicates that this star is actually a SX Phe star. In addition, the
$\log g= 3.8$ obtained is in good agreement with the spectroscopic
study referenced in \citet{2015MNRAS.447.3264S}, but we note that our model
has a higher effective temperature (around $900$ K). This could be
due to the fact that we include in our procedures only the dominant
frequencies, and not all the detected ones, as it was done in
\citet{2015MNRAS.447.3264S}. Despite these differences, our best-fit
model for KIC 9244992 is also at the end of the MS stage.

\begin{table}[!ht]
  \centering
  \caption{Best-fit model for KIC  9244992.}
  \begin{tabular}{lcc}
    \hline\hline\noalign{\smallskip}
                 & Procedure 1 \\ \hline\noalign{\smallskip}

    $M_{\star} [M_{\rm \sun}]$ &   2.10  \\ 
    $Z$                      &    0.02   \\
    $f$                      &     0      \\
    $T_{\rm eff}$ [K]         &    8150    \\
    $\log g$            &    3.89   \\
    $R_{\star} [R_{\rm \sun}]$ &    2.7 \\
    Age[$10^6$ yr]        &   690.94  \\
    $L_{\star} [L_{\rm \sun}]$ &  31.25 \\
    $\overline{\Delta \Pi}$ [sec]&    2271.8  \\
           \hline
  \end{tabular}
  \label{modelossaioprocedures}
\end{table}

\subsection{CoRoT 105733033} %chapellier  

According to the measurements provided by the EXODAT database, CoRoT
105733033 as a A5V spectral type with magnitude $V= 12.8$ and
effective temperature of $8000$ K. This star is a good example of
hybrid pulsators since it shows $g$ and $p$ modes in two clearly
distinct frequency domains. \citet{2012A&A...540A.117C} divided the
CoRoT 105733033 periods into two domains: the $\gamma$ Dor domain with
periods in the range $[21600-345600]$ sec and the $\delta$ Sct domain
with periods in the range $[1362.77-8554.45]$ sec,
with the largest-amplitude mode in the $\delta$ Sct domain with a period
$9816.08$ sec, and a mean period spacing of $g$ modes
of $2655.936$ sec. To perform our calculations we used a $g$-mode
period range of $[59961.6-137635.2]$ sec.
Table \ref{tabladatoschap} summarizes these stellar parameters and the
observed pulsation-period ranges for CoRoT 105733033.

\begin{table}[!ht]
  \centering
  \caption{Observational data for CoRoT 105733033.}
  \begin{tabular}{lc}
    \hline\hline\noalign{\smallskip}
          $T_{\rm eff}$ [K]  &  8000  \\ 
Period range of $\delta$ Sct domain [sec] & $[1362.77-8554.45]$ \\
Period range of $\gamma$ Dor domain [sec] & $[59961.6-137635.2]$ \\
 $\overline{\Delta \Pi}$ [sec]&  2655.93   \\
Period of the largest-amplitude mode [sec]&   6816.08  \\            
           \hline\hline
  \end{tabular}
  \label{tabladatoschap}
\end{table}

\begin{figure}[h!] 
\begin{center}
\includegraphics[clip,width=9 cm]{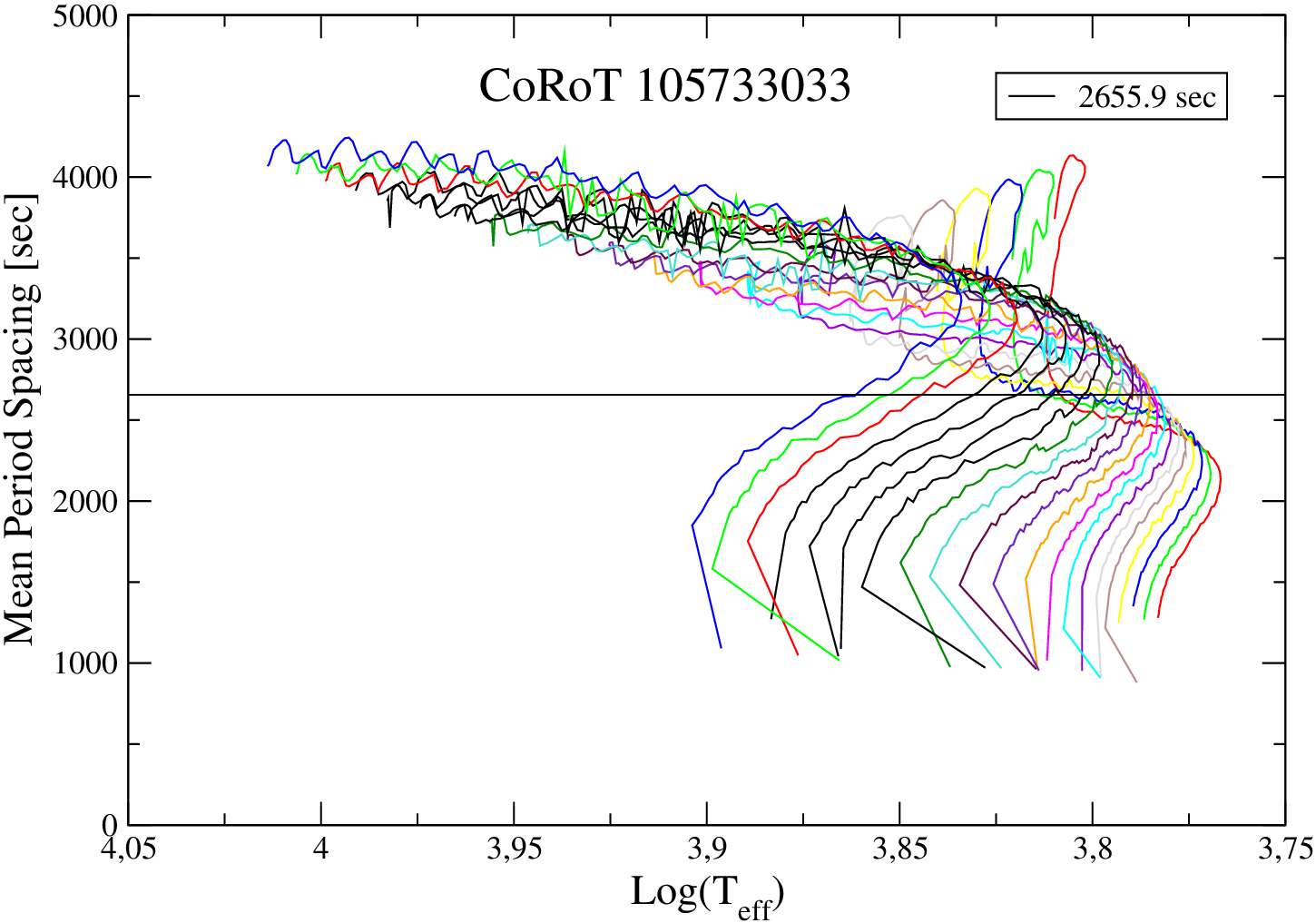} 
\caption{Mean period spacing vs effective temperature for
  models with $Z= 0.015$ and $f= 0.03$. The straight line is the observed
  mean period spacing for $g$ modes of CoRoT 105733033. }
\label{figureChapDP} 
\end{center}
\end{figure}

\begin{figure}[h!] 
\begin{center}
\includegraphics[clip,width=9 cm]{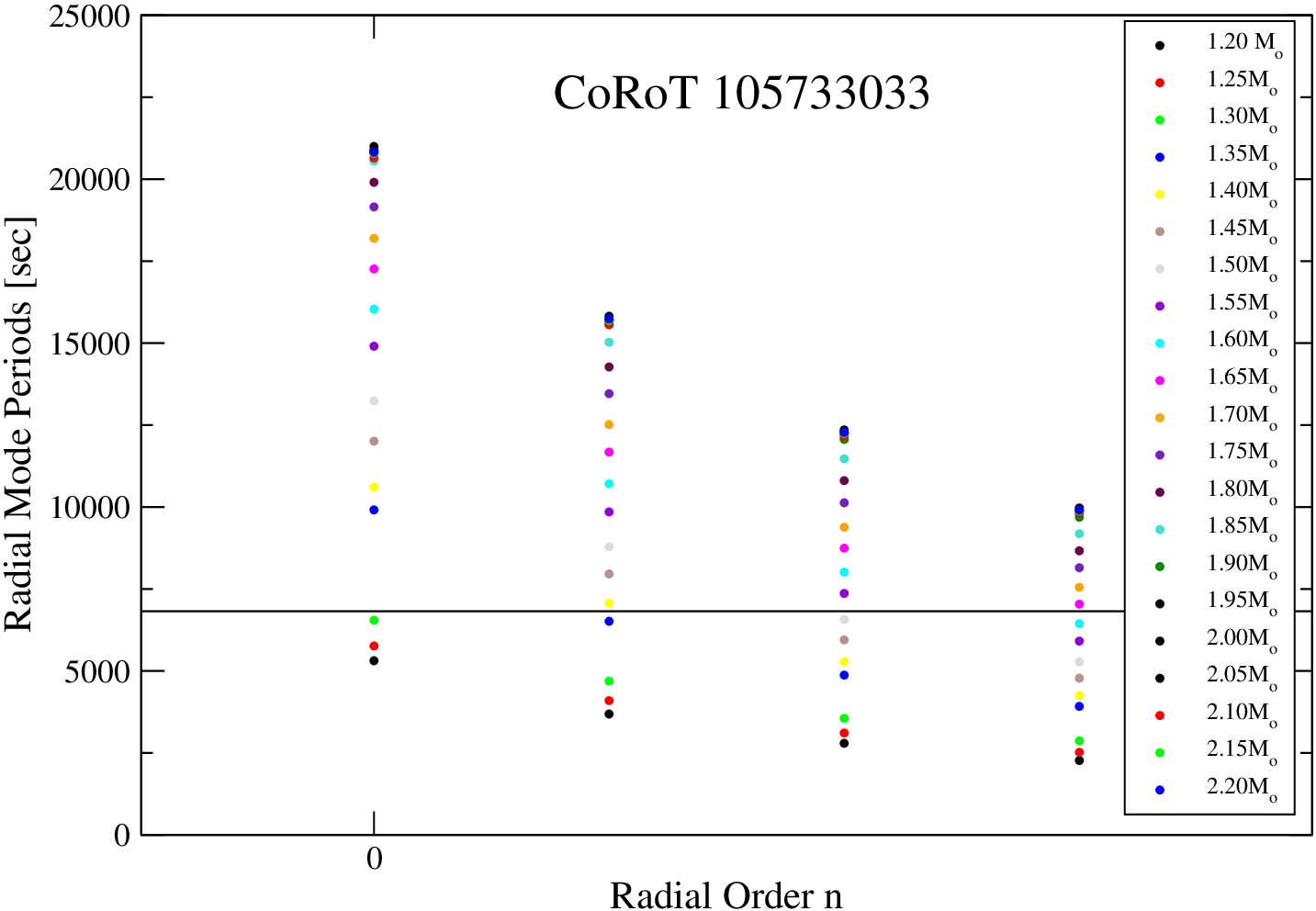} 
\caption{Radial mode periods vs radial order $n$ for models selected
  in the previous step with $Z= 0.015$ and $f= 0.03$. }
\label{figurechapradial} 
\end{center}
\end{figure}

In Figure \ref{figureChapDP} we display the variation of the mean
period spacing in the $g$-mode range with the effective temperature
for models with $Z= 0.015$ and $f= 0.03$. We calculated
$\overline{\Delta \Pi_n}$ using modes with $\ell= 1$. Next, we performed
the \textit{Step 3} of \textbf{Procedure 1}. Figure
\ref{figurechapradial} shows the radial modes for selected models with
$Z= 0.015$ and $f= 0.03$.

The complete list of the detected frequencies can be found in Table 1
of \citet{2012A&A...540A.117C}. There are 444 frequencies classified
as $\delta$ Sct-type or $\gamma$ Dor-type shown in the aforementioned
table. We performed our calculation dismissing all kinds of
combination frequencies. Since none of the $p$- or $g$-mode periods
have an assigned harmonic degree, we calculated $\chi^p$ and $\chi^g$
searching for the best period fit among modes with $\ell= 1, 2$ and $3$.
As it was done for HD 49434, the largest-amplitude mode
(classified as F in Table 1 of \citet{2012A&A...540A.117C}) was
dismissed in the calculation of $\chi^p$ in \textbf{Procedure 1}, but included in the calculation of $\chi^{rp}$ in \textbf{Procedure 2}.

In Table \ref{modeloschapprocedures} we show the main characteristics
of the asteroseismological models selected for CoRoT 105733033. So
far, no asteroseismological model has been proposed for this star, and their
physical characteristics are quite uncertain. Here, we present the
first asteroseismic model of this star, with the parameters displayed
in Table \ref{modeloschapprocedures}. As it was mentioned, we performed \textbf{two} different procedures and we obtained two different models depicted in
Fig. \ref{figurerhmodelosfinales}, both reaching the
TAMS. Again, we calculated $\chi^g$ in order to add it in $F_1$ for \textbf{Procedure 1} and we obtained the
same model (the one with $1.75 M_{\sun}$) and another different model
from \textbf{Procedure 2} (the one with $1.85 M_{\sun}$). This
means that the best-fit model selected persists when we consider a
period-to-period fit of $g$-modes, but changes when we include the
possibility of having radial modes between the frequencies detected in
the $\delta$ Sct domain. Anyway, both models are before the TAMS, and have the same overshooting parameter and similar
effective temperatures and surface gravities.

\begin{table}[!ht]
  \centering
  \caption{Best-fit models for CoRoT 105733033.}
  \begin{tabular}{lcc}
    \hline\hline\noalign{\smallskip}
                 & \textbf{Procedures 1}  & \textbf{Procedure 2}\\ 
\hline\noalign{\smallskip}
    $M_{\star} [M_{\rm \sun}]$ &  1.75 & 1.85 \\ 
    $Z$                      &    0.015 &  0.01 \\
    $f$                      &     0.03 &  0.03   \\
    $T_{\rm eff}$ [K]         &    6169  & 6537 \\
    $\log g$                 &   3.5 &  3.45  \\
    $R_{\star} [R_{\rm \sun}]$ & 3.85 & 4.19\\
    Age[$10^6$ yr]           & 1628.4& 1272.79 \\
    $L_{\star} [L_{\rm \sun}]$ &  19.77&  30.44\\
    $\overline{\Delta \Pi}$ [sec]&   2646 &  2682.36 \\
           \hline
  \end{tabular}
  \label{modeloschapprocedures}
\end{table}

\subsection{CoRoT 100866999} %chapellier binaria

This star is an eclipsing binary system, with a pulsating primary star
compatible with an A7-F0 spectral type and the secondary star with a
G5-K0 spectral type \citep{2013A&A...556A..87C}. From the eclipsing
curve fit, these authors found a stellar mass of $(1.8 \pm 0.2) M_{\rm
  \sun}$, a radius of $(1.9 \pm 0.2) R_{\rm \sun}$, a surface gravity
of $\log g= 4.1 \pm 0.1$ and an effective temperature  of $(7300 \pm
250)$ K for the primary star. For the secondary, they found a stellar
mass of $(1.1 \pm 0.2) M_{\rm \sun}$, a radius of $(0.9 \pm 0.2)
R_{\rm \sun}$, a surface gravity of $\log g =4.6 \pm 0.1$ and an
effective temperature of $(5400 \pm 430)$ K. The pulsating primary
star has two well separated $\delta$ Sct and $\gamma$ Dor period domains.
The $g$-mode period range of this star is $[23736-288000]$ sec and the
$p$-mode periods are in the range $[2544.16-5927.14]$ sec. CoRoT
100866999 has a mean period spacing of $g$ modes equal to $3017.952$
sec and presents the largest-amplitude mode in the $\delta$ Sct period
range at a period of 5088.24 sec. Table \ref{tabladatoschapbin}
summarizes these stellar parameters and the observed pulsation period
ranges for CoRoT 100866999.

\begin{table}[!ht]
  \centering
  \caption{Observational data for CoRoT 100866999.}
  \begin{tabular}{lc}
    \hline\hline\noalign{\smallskip}
          $T_{\rm eff}$ [K] &  $7300 \pm 250$ \\
   $\log g$  &  $4.1 \pm 0.1$  \\
Period range of $\delta$ Sct domain [sec] & $[2544.16-5927.14]$ \\
Period range of $\gamma$ Dor domain [sec] & $[23736-288000]$  \\ 
$\overline{\Delta \Pi}$ [sec]&  3017.952 \\
Period of the largest-amplitude mode [sec]& 5088.249 \\
$M_{\star} [M_{\rm \sun}]$ &  $1.8 \pm 0.2$\\
$R_{\star} [R_{\rm \sun}]$ & $1.9 \pm 0.2$   \\
           \hline\hline
  \end{tabular}
  \label{tabladatoschapbin}
\end{table}

\begin{figure}[h!] 
\begin{center}
\includegraphics[clip,width=9 cm]{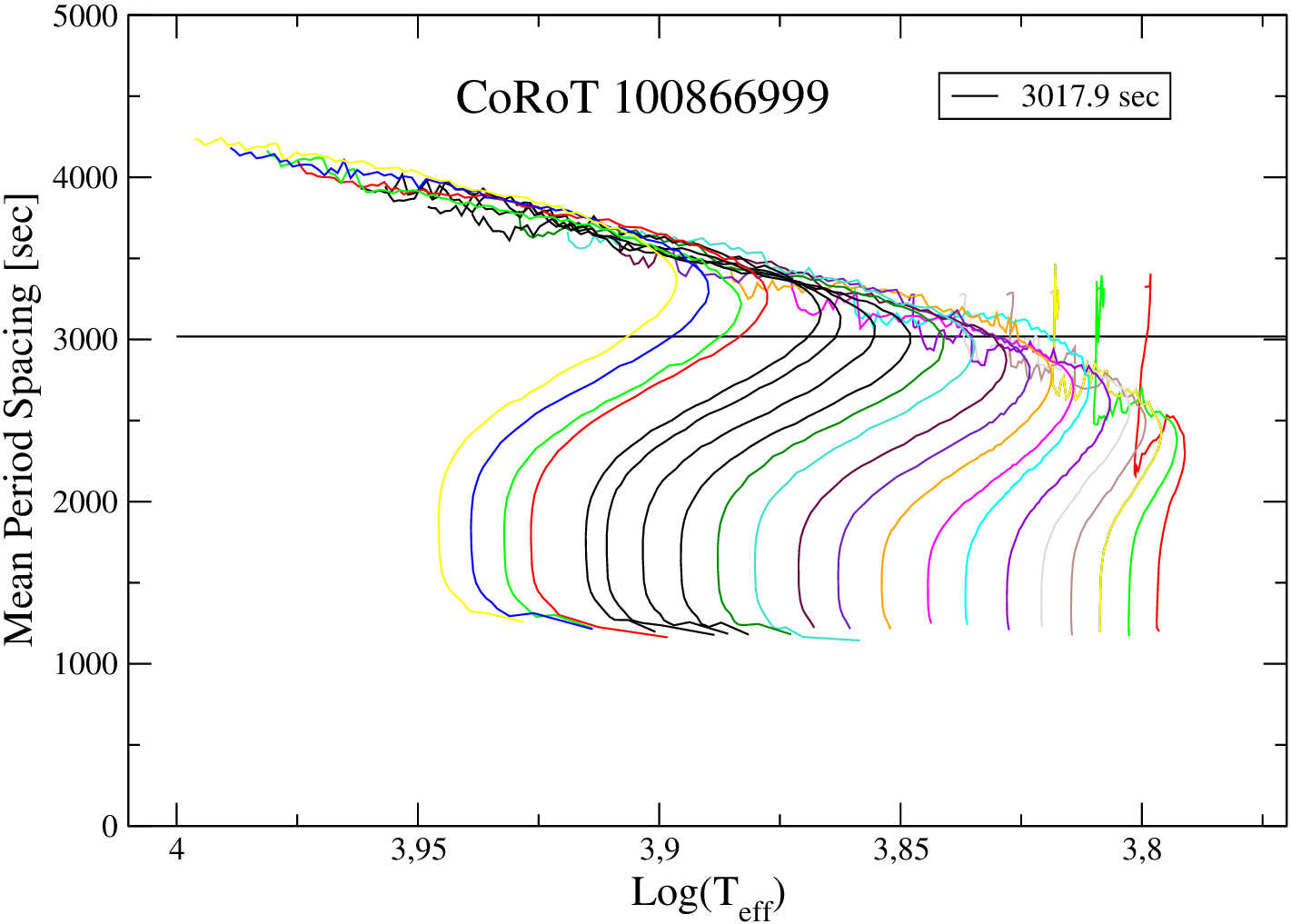} 
\caption{Mean period spacing of $g$ modes vs effective temperature for
  models with $Z= 0.02$ and $f= 0$. The horizontal straight line is
  the observed mean period spacing for $g$-modes of CoRoT 100866999.}
\label{figureChapbinDP} 
\end{center}
\end{figure}

We calculated $\overline{\Delta \Pi_n}$ using modes with $\ell= 1$,
since $\Delta \Pi$ separation is compatible with $\ell= 1$ $g$ modes, as
mentioned by \citet{2013A&A...556A..87C}. Figure \ref{figureChapbinDP}
depicts the calculated mean period spacing for models with $Z= 0.02$
and $f= 0$.

\begin{figure}[h!] 
\begin{center}
\includegraphics[clip,width=9 cm]{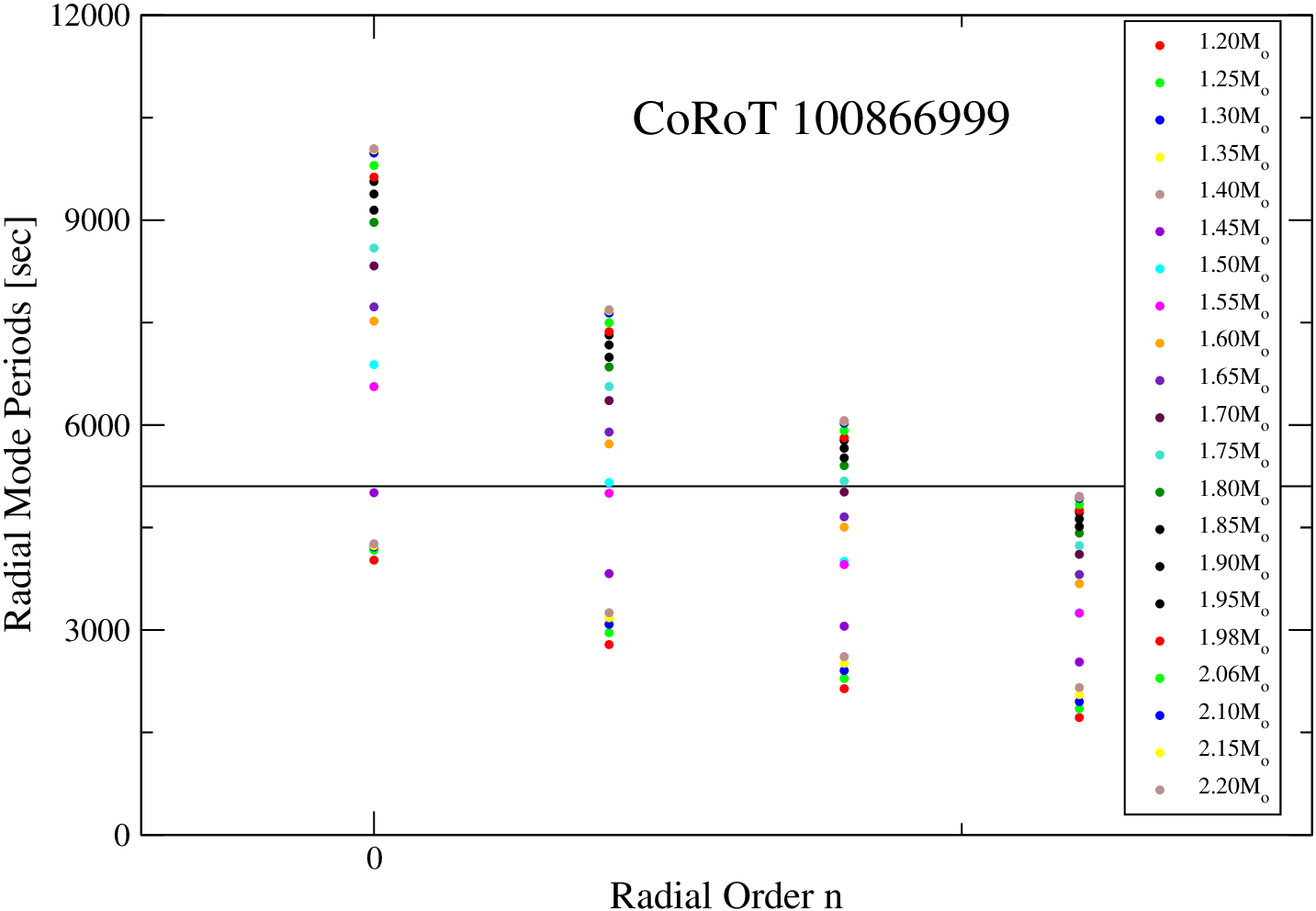} 
\caption{Radial mode periods vs radial order $n$ for models
  selected in the previous step with $Z= 0.02$ and $f= 0$,
  corresponding to CoRoT 100866999. }
\label{figurechapbinradial} 
\end{center}
\end{figure}

For the \textit{Step 2} of \textbf{Procedure 1} we adopted the
largest-amplitude mode frequency as the radial fundamental mode,
i.e. 5088.249 sec. The radial-mode periods for the models selected
in the previous step with $Z= 0.02$ and $f= 0$ are depicted in Figure
\ref{figurechapbinradial}. As it was done previously, we only selected
models whose fundamental or an overtone radial mode is 100 sec or less
away from the largest-amplitude mode detected in CoRoT 100866999.

Again, we calculated $\chi^p$ and $\chi^g$ in the same way as for HD
49434 and CoRoT 105733033, since there are no harmonic degree
classification for none of the detected mode periods. This fact
allowed us to perform \textbf{Procedure 2}. In Table
\ref{modeloschapbinprocedures}, we display the
characteristics of the models selected for CoRoT 100866999.

\begin{table}[!ht]
  \centering
  \caption{Best-fit model for CoRoT 100866999.}
  \begin{tabular}{lcc}
    \hline\hline\noalign{\smallskip}

                 & \textbf{Procedures 1} & \textbf{Procedure 2}\\ \hline\noalign{\smallskip}
    $M_{\star} [M_{\rm \sun}]$ &  1.55 & 2.10 \\ 
    $Z$             &   0.02 &  0.02 \\
    $f$              &     0 &  0   \\
    $T_{\rm eff}$ [K]     &    6778  & 7726 \\
    $\log g$            & 4.04 &  3.84  \\
    $R_{\star} [R_{\rm \sun}]$ & 1.94 &  2.86\\
    Age[$10^6$yr]        &  1298.32 & 682.90 \\
    $L_{\star} [L_{\rm \sun}]$ &  7.49 & 28.23\\
    $\overline{\Delta \Pi}$ [sec]&  3020.89&  3004.10 \\
               \hline
  \end{tabular}
  \label{modeloschapbinprocedures}
\end{table}

We obtained one model from \textbf{Procedure 1} with and without considering $\chi^g$ in $F_1$ and a
different model from \textbf{Procedure 2}. The one obtained from
\textbf{Procedure 1} with $1.55 M_{\sun}$ is
located before the ``knee'' of the MS. The other one,
obtained with \textbf{Procedure 2}, has $2.10 M_{\sun}$ and it is
near the TAMS. Both models have $Z= 0.02$. These values are close to
those obtained in \citet{2013A&A...556A..87C} from the eclipsing curve
fit for the primary star ($1.8 \pm 0.2 M_{\sun}$). 

With the aim of quantify our model selections, we show in Table \ref{cuantifica} the differences between the observed and the computed period spacing of g modes, $\delta(\Delta \Pi)$; the observed and computed period of the highest-amplitude radial mode, $\delta(\Pi^r)$; and the quantities $\sigma^p$, $\sigma^g$ and $\sigma^{rp}$. The quantities $\sigma^p$, $\sigma^g$ and $\sigma^{rp}$ are defined as:

\begin{equation}
\sigma^j= \frac{1}{N}\sum _{i=1}^N \vert \Pi^j_o-\Pi^j_c \vert_i \,\,, \quad j=p,g,rp 
\end{equation}

\noindent Note that $\sigma^{rp}$ has been computed only for target stars with no clear mode identification, as \textbf{Procedure 2} requires.
 
\begin{table}[!ht]
  \centering
  \caption{The O-C differences of each method and target star. All numbers are seconds. The second arrow for KIC 11145123 has also been computed for the model obtained with \textbf{Procedure 1} but with the period-to-period fit of the individual $g$ modes.}
  \begin{tabular}{lcccccc}
    \hline\hline\noalign{\smallskip}

          Star     & $\!\!\!\!\!\!$Method$\!\!\!\!$ &$\!\!$ $\delta(\Delta \Pi)$ & $\delta(\Pi^r)$& $\sigma_p$& $\sigma_g$ & $\sigma_{rp}$ \\ \hline\hline\noalign{\smallskip}

    HD      & $\!\!\!\!\!\!$1$\!\!\!\!$ &$\!\!$15.05                    &    26.41            &   142.96 &     51.35  &     127.75   \\
      49434 &$\!\!\!\!\!\!$ 2$\!\!\!\!$&$\!\!$ 3.48                    &    751.07           &   86.56  &    278.50  &     61.16    \\ \hline
    KIC     &$\!\!\!\!\!\!$1$\!\!\!\!$&$\!\!$11.57                &     4.78             & 69.53  &    765.48  &        -       \\
    11145123 &$\!\!\!\!\!\!$1 ($\chi^g$)$\!\!\!\!$&$\!\!$  1.73                &     7.04           &  97.99   &    702.09  &        -      \\ \hline
    KIC       &$\!\!\!\!\!\!$1$\!\!\!\!$& $\!\!$     8.2                 &     5.84           &   189.27 &   592.73  &     -       \\ 
    9244992   &  &                    &                     &           &         &              \\\hline
     CoRoT      &$\!\!\!\!\!\!$ 1$\!\!\!\!$& $\!\!$9.93                   &    22.67           &    57.79 &    203.13   &     48.72     \\
       105733033 &$\!\!\!\!\!\!$2$\!\!\!\!$&$\!\!$ 26.43                    &    338.73          &    54.81 &    204.87 &     41.86        \\ \hline
   CoRoT    &$\!\!\!\!\!\!$1$\!\!\!\!$&$\!\!$ 2.93                    &     85.39          &    25.82 &   238.07   &  26.69            \\
   100866999 &$\!\!\!\!\!\!$2$\!\!\!\!$& $\!\!$    13.85                     &    165.65         &   47.72    & 240.32    &   16.64            \\  
       \hline\hline
  \end{tabular}
  \label{cuantifica}
\end{table}

\section{Summary and conclusions}
\label{conclusions}

In this work, we have presented a detailed asteroseismic study of five
hybrid $\delta$ Scuti-$\gamma$ Doradus pulsating stars, aiming at
derive their fundamental stellar parameters. To this end we built a
huge grid of stellar models, covering the evolution of low-mass stars
from the ZAMS to the TAMS, varying the stellar mass, the metallicity
and the amount of core overshooting (see Sect. \ref{Modelling}). We
employed the observational data of the detected periods reported in
\citet{2014MNRAS.444..102K} for KIC 11145123,
\citet{2015MNRAS.447.3264S} for KIC 9244992,
\citet{2015MNRAS.447.2970B} for HD 49434, \citet{2012A&A...540A.117C}
for CoRoT 105733033, and \citet{2013A&A...556A..87C} for CoRoT
100866999. We were able to obtain the fundamental parameters of the
target stars by performing two different procedures, which fully
exploit the simultaneous presence of $p$ and $g$ modes (and also
presumably radial modes) in this kind of pulsating stars. 

For \textbf{Procedure 1} we used three constraints to find the
best-fit seismological models: 1) the mean period spacing of
high-order $g$ modes; 2) the largest amplitude mode in the $\delta$
Sct period domain, to which we associate a radial mode; and 3) a
period-to-period fit of the individual $p$ modes. Besides, in this case we explore the incidence of adding a period-to-period fit of $g$ modes in the selection of the best fit model. Finally in \textbf{Procedure 2} we used again the mean period spacing of $g$ modes and, in addition, a period-to-period fit between the frequencies detected in the $\delta$ Sct domain and those
calculated for models with $p$-mode periods and also (simultaneously)
radial-mode periods. It is worth mentioning that these procedures do
not depend on the reported spectroscopic information of the target stars,
e.g., the effective temperature. 

Below, we summarize the results obtained for each star:

\begin{figure}[h!] 
\begin{center}
\includegraphics[clip,width=9 cm]{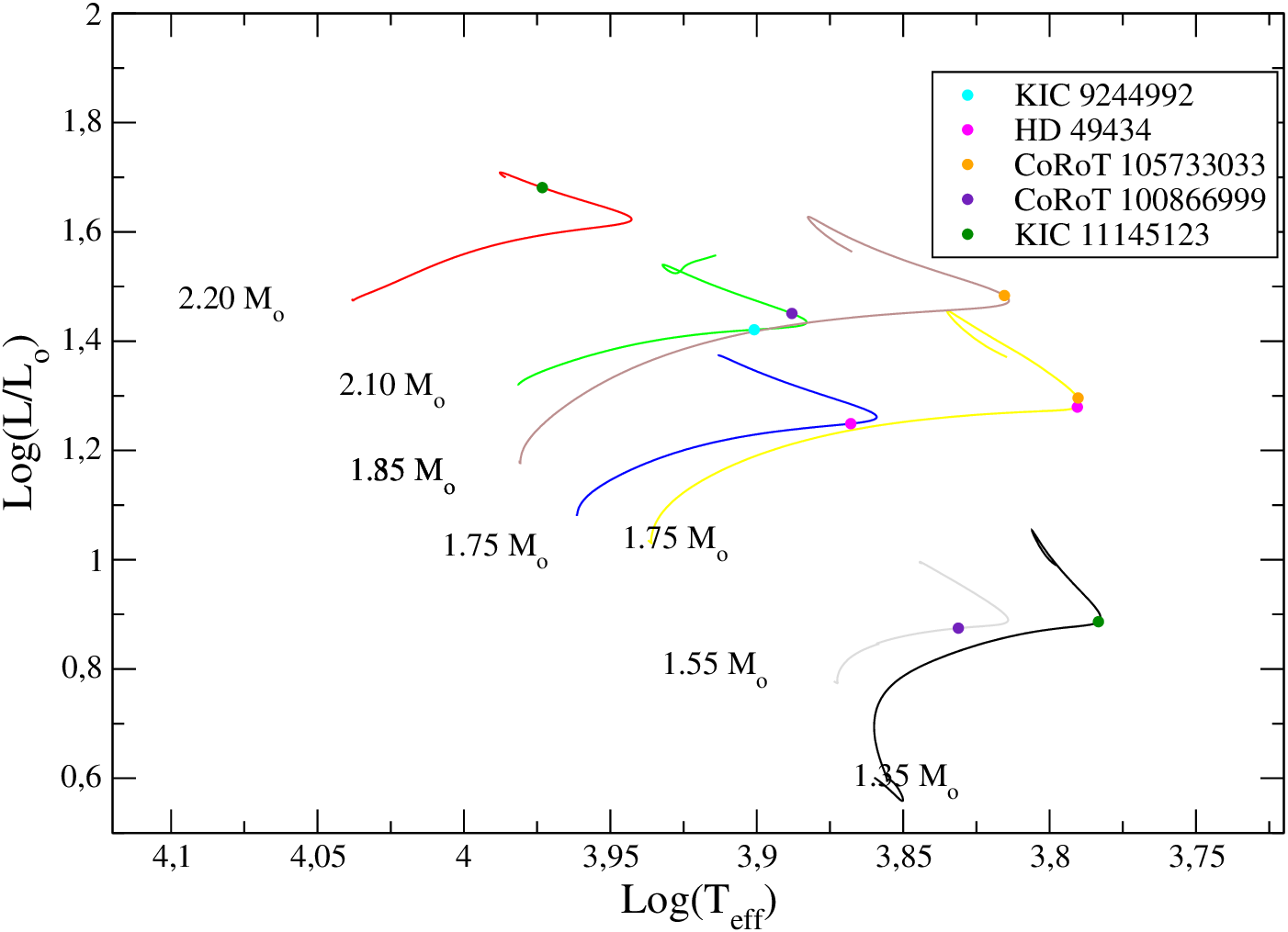} 
\caption{HR diagram showing the asteroseismological models
  found and the respective evolutionary tracks for each target star.}
\label{figurerhmodelosfinales} 
\end{center}
\end{figure} 

\begin{itemize}

\item {\bf KIC 11145123}: Two different seismological models were
  obtained from the \textbf{Procedures 1} with and without including $\chi^g$ in $F_1$(see Table
  \ref{modeloskurtzprocedures}). This means that individual $g$-mode
  period fits play an important role in the modeling of this
  star. Comparing our models with those obtained in
  \citet{2014MNRAS.444..102K}, we note some differences in the mass and
  the metallicity, possibly due to the fact that we did not considered
  atomic diffusion in our simulations ---the Brunt-V\"ais\"al\"a
  frequency is modified by this physical process--- and also that we
  neglected rotation in our pulsation modeling. Nevertheless, our best-fit
  models are both in the TAMS overall contraction phase, in good agreement with
  those obtained in \citet{2014MNRAS.444..102K}, and also have a
  metallicity  consistent with the hypothesis that this star could be
  a SX Phe star. Note that none of both models reproduce well the
  reported effective temperature for this star.

\item {\bf KIC 9244992}: The characteristics of the model obtained
  for this star are shown in Table \ref{modelossaioprocedures}. In
  this case, we obtained the same best-fit model with and without including $\chi^g$ in \textbf{Procedures 1}. This means that including a
  $g$-mode period-to-period fit does
  not affect the selection of the model. Our $2.1M_{\sun}$
  best-fit model is more massive that the one proposed in
  \citet{2015MNRAS.447.3264S} of $1.45M_{\sun}$. Also, our model
  does not have overshooting in the core and has higher metallicity
  ($Z= 0.02$). Thus, our results do not indicates that this star is
  actually a SX Phe star. In addition, the $\log g= 3.8$ obtained is in
  good agreement with the spectroscopic study cited in
  \citet{2015MNRAS.447.3264S}, but we note that our model has an
  higher effective temperature. Despite these differences, our best fit
  model for KIC 9244992 is also at the end of the MS stage.

\item {\bf HD 49434}: We perform \textbf{Procedure 2} for this star
  since the mode classification is not conclusive and therefore the
  existence of radial modes in the $\delta$ Sct region cannot be
  discarded. We obtained one model with $Z= 0.01$ and $f= 0.01$ from
  \textbf{Procedure 1}, and another model with
  $Z= 0.015$ and $f= 0.03$ from \textbf{Procedure 2}. Both models have
  the same mass, $1.75 M_{\sun}$. One of them has $f= 0.03$ and is
  \textbf{near} the TAMS, and the other one has $f= 0.01$ and is
  before the evolutionary ``knee'' where the overall contraction phase begins, as can
  be observed in Fig. \ref{figurerhmodelosfinales}. One of the main
  characteristics
  of this star is that it is a rapid rotator, and so it not shows
  a clear gap between the $\delta$ Sct and $\gamma$ Dor pulsation spectra
  regions. It is possible, as it is mentioned in
  \citet{2015MNRAS.447.2970B}, that the absence of the gap is due
  precisely to rotational splitting of high-degree $p$ modes. If this
  is the case, it is necessary a correct mode identification since our
  methodology  strongly depends on this. On the other hand it is worth
  mentioning that adding a period-to-period fit for the $g$-modes does
  not affect the selection of the best fit model since we obtained the
  same models including $\chi^g$ in \textbf{Procedure 1}. The
  selection of another model when we include the possibility of having
  radial modes in the $\delta$ Sct period domain shows and reinforces
  the need of having a correct mode classification.
%%ver acá lo que dicen estos autores sobre los modos radiales, no lo entendí.

\item {\bf CoRoT 105733033}: One of the remarkable characteristics
  of this star is the richness of its pulsational spectra. As it was
  mentioned before, it is possible to observe a clear distinction
  between low- and high-frequency regions in this star, which may be
  the consequence of a quite low angular rotation
  \citep{2012A&A...540A.117C}. More spectroscopic data is required to
  confirm this hypothesis.  So far, no asteroseismological model
  has been proposed
  for this star, and their physical characteristics are uncertain. In
  this paper, we present the first asteroseismic model (see
  Table \ref{modeloschapprocedures}). As
  it was mentioned, we performed three different procedures and we
  obtained two different models, both at the overall contraction phase (Fig. \ref{figurerhmodelosfinales}).
  Again, from \textbf{Procedure 1} when we include $\chi^g$ we obtained the same model (one with $1.75 M_{\sun}$) and another different
  model from \textbf{Procedure 2} (the one with $1.85 M_{\sun}$).
  This means that the best fit model selected persists
  when we consider a period-to-period fit of $g$-modes, but changes
  when we include the possibility of having radial modes among the
  frequencies detected in the $\delta$ Sct domain. Anyway, both models
  are at the overall contraction phase, and have the same overshooting
  parameter and similar effective temperatures and surface
  gravities.

\item {\bf CoRoT 100866999}: We obtained one model from
\textbf{Procedure 1} and a different model from
  \textbf{Procedure 2}. The one obtained with \textbf{Procedure 1} has $1.55 M_{\sun}$ and is located before the evolutionary 
  ``knee'' of the MS. The other one, obtained with
  \textbf{Procedure 2}, has $2.10 M_{\sun}$ and is on the overall contraction phase. Both models have $Z= 0.02$. Comparing these mass values with
  the ones obtained in \citet{2013A&A...556A..87C} from the eclipsing
  curve fit, we can see that ours masses are close to those calculated
  for the primary star ($1.8 \pm 0.2 M_{\sun}$).

\end{itemize}

In summary, we have obtained for the first time reliable
asteroseismological models representative of five $\delta$
Sct-$\gamma$ Dor hybrid stars by means of grid-based modeling. These
asteroseismological models result from different criteria of model
selection, in which we take full advantage of the richness of periods
that characterizes the pulsation spectra of this kind of stars.
For four out the five stars analyzed, we have obtained the same
asteroseismological model from \textbf{Procedure 1} including or not a period-to-period fit of the $g$ modes. In the cases when it was possible to apply \textbf{Procedure 2}, we
obtained a different model from this approach. It is worth of notice that the true seismic model for a given target star must reproduce not only observed frequencies and regularities in the frequency spectra, but also frequency ranges of observed oscillations as ranges of pulsationally unstable radial and nonradial modes. We considered only adiabatic oscillations in our approach, and a detailed stability analysis of oscillations, which is beyond the scope of the present work, will be addressed in a future paper. Clearly, more theoretical work in the frame of this issue (like nonadiabatic stability computations), and also in other topics, for instance, the inclusion of the effects of rotation on the pulsation periods and substantial improvement of mode identification, will help us to break the degeneracy of the asteroseismological solutions.

\begin{acknowledgements}
We wish to thank our anonymous referee for the constructive
comments and suggestions that greatly improved the original version of
the paper. The complete grid of stellar models used in this paper was computed with the calculus cluster of the ``Instituto de F\'isica de L\'iquidos y Sistemas Biol\'ogicos'' which belong to the ``National System for High Performance Computing'' (SNCAD) of the Ministry of Science, Technology and Productive Innovation-Argentina. Part of this work was
supported by AGENCIA through the Programa de Modernizaci\'on
Tecnol\'gica BID 1728/OC-AR, and by the PIP 112-200801-00940 grant
from CONICET. This research has made use of NASA Astrophysics Data
System.    
\end{acknowledgements}

\bibliographystyle{aa} % style aa.bst
\bibliography{paper-hybrids} % your references Yourfile.bib

\end{document}